\documentclass[11pt,english]{article}
\usepackage[T1]{fontenc}
\usepackage[latin9]{inputenc}
\usepackage{geometry}
\usepackage{float}
\usepackage{mathrsfs}
\usepackage{amsmath}
\usepackage{amssymb}
\usepackage{graphicx}
\pdfoutput=1 
\usepackage{subfig}
\usepackage{babel}
\begin{document}

\title{Sampling microcanonical measures of the 2D Euler equations through
Creutz's algorithm: a phase transition from disorder to order when
energy is increased }

\author{Max Potters, Timothee Vaillant and Freddy Bouchet}
\maketitle
\begin{abstract}
The 2D Euler equations is the basic example of fluid models for which
a microcanical measure can be constructed from first principles. This
measure is defined through finite-dimensional approximations and a
limiting procedure. Creutz's algorithm is a microcanonical generalization
of the Metropolis-Hasting algorithm (to sample Gibbs measures, in
the canonical ensemble). We prove that Creutz's algorithm can sample
finite-dimensional approximations of the 2D Euler microcanonical measures
(incorporating fixed energy and other invariants). This is essential
as microcanonical and canonical measures are known to be inequivalent
at some values of energy and vorticity distribution. Creutz's algorithm
is used to check predictions from the mean-field statistical mechanics
theory of the 2D Euler equations (the Robert-Sommeria-Miller theory).
We found full agreement with theory. Three different ways to compute
the temperature give consistent results. Using Creutz's algorithm,
a first-order phase transition never observed previously, and a situation
of statistical ensemble inequivalence are found and studied. Strikingly,
and contrasting usual statistical mechanics interpretations, this
phase transition appears from a disordered phase to an ordered phase
(with less symmetries) when energy is increased. We explain this paradox. 
\end{abstract}

\section{Introduction}

Two-dimensional and geophysical flows are highly turbulent, yet embody
large-scale coherent structures such as ocean rings, jets, and large-scale
vortices. Understanding how these structures appear and predicting
their shape is a major theoretical challenge. The statistical mechanics
approach to geophysical flows is a powerful complement to more conventional
theoretical and numerical methods (see Ref. \protect\cite{bouchet_venaille2}
for a recent review, or Refs. \protect\cite{Lim,Majda_Wang_Book_Geophysique_Stat, Lim_Ding_Nebus_2010}
for related approaches). In the inertial limit statistical equilibria
describe, with only a few thermodynamical parameters, the main natural
attractors of the dynamics. 

Recent studies in quasi-geostrophic models provide encouraging results:
a model of the Great Red Spot of Jupiter \protect\cite{bouchet_sommeria1},
and explanations of several different phenomena: the drift properties
of ocean rings \protect\cite{bouchet_venaille1}, the inertial structure of
mid-basin eastward jets \protect\cite{bouchet_venaille1}, bistability in
complex turbulent flows \protect\cite{bouchet_simonnet1}, and so on. 

Generalization to more comprehensive hydrodynamical models, which
can simulate gravity-wave dynamics and enable energy transfer through
wave motion, would be interesting. Both of the aforementioned processes
are essential in understanding the geophysical flow energy balance.
However, due to difficulties in essential theoretical parts of the
statistical mechanics approach, previous methods describing statistical
equilibria were up to now limited to the use of quasi-geostrophic,
or simpler, models. In order to study the statistical mechanics of
those models, it would be useful to be able to rely on numerical sampling
of their microcanonical measures. \\

The 2D Euler equations can be formulated in terms of the vorticity
field. Points in the vorticity field are coupled through a long-range
interaction. In contrast with traditional systems, long-range interaction
systems are well known to show generic inequivalence between microcanonical
and canonical ensembles (\protect\cite{Bouchet_Barre2005,Bouchet_Gupta_Mukamel2010,Campa_Dauxois_Ruffo2009,Chavanis_2006IJMPB_Revue_Auto_Gravitant,EllisHavenTurkington:2000_Inequivalence},
see Ref. \protect\cite{Bouchet_Barre2005} for a classification). The microcanonical
ensemble is richer than the canonical one, as the canonical equilibrium
states form a subset of the microcanonical equilibrium states \protect\cite{EllisHavenTurkington:2000_Inequivalence,Bouchet_PhysicaD_2008}.
For these systems it is thus essential to be able to sample microcanonical
measures instead of canonical ones. 

Creutz's algorithm \protect\cite{creutz} is a Monte-Carlo approach used to
sample microcanonical measures of discrete spin systems \protect\cite{Mukamel-Ruffo_Schreiber_2005PhRvL,Gupta_Mukamel2010}.
Whereas the Metropolis-Hasting algorithm samples Gibbs measures (canonical
ensemble), the Creutz algorithm imposes an energy constrain and thus
samples the microcanonical measure (microcanonical ensemble). Section
\protect\ref{sec:Creutz-algorithm} gives a proof of detailed balance and
convergence to the microcanonical measure for Creutz's algorithm. Appendix A defines the classical concepts useful for the discussion of detailed balance. Appendix B also describes some improper interpretation of the algorithm
that may lead to wrong results. 

The main aim of this paper is to discuss the first generalization
of Creutz's algorithm to hydrodynamical systems. The main novelty
is the ability to deal with the statistical mechanics of fields rather
than discrete variables. For this first work, we consider the 2D Euler
equations and precisely define the microcanonical measures through
finite-dimensional approximations and a limiting procedure. The generalization
to the Quasi-Geostrophic model or the Vlasov equation, the microcanonical
measure definition and their sampling through Creutz's algorithm would
be straightforward. The method is extremely robust and could also
be easily generalized to more complex models like the 3D axisymmetric
Euler equations or the Shallow Water model. \\

The 2D Euler equations and related models have an infinite number
of conserved quantities, called Casimirs. Michel and Robert \protect\cite{robert_michel} have first discussed the use of large deviation theory as a justification of mean field variational problems for the microcanonical measure. Later on, Ellis and collaborators \protect\cite{Boucher_Ellis_Turkington_2000_JSP}
have defined approximations of  equilibrium measures,  through a description of the fields over a lattice, where $N^{2}$ is the number of degrees of freedom (lattice site). In this work, Ellis and Turkington have treated the energy constraint  microcanonically and the Casimir constraints canonically. They have proven that these approximate measures verify a large deviation principle, where  $N^{2}$ is the large deviation rate.  In this paper, we study a $N^{2}$-degrees-of-freedom discretized approximation of
equilibrium measures (following Ellis and collaborators), but treating all constraints microcanonically (as did Robert and Michel). This slightly different presentation is an improvement, as proceeding through discretization provides a clear and straightforward
definition of the microcanonical measures, and as the set
of microcanonical equilibrium states includes the set of canonical equilibrium states (see the beginning of the introduction). The limit is then an invariant
measure of the 2D Euler equations \protect\cite{Bouchet_Corvellec2010}.

As was already clear in previous works \protect\cite{robert_michel, Boucher_Ellis_Turkington_2000_JSP} (please see a detailed discussion in Ref. \protect\cite{bouchet_venaille2}),
the 2D Euler equations show mean-field behavior. It is therefore natural
to define macrostates through coarse-graining of microstates. The
most probable macrostate maximizes an entropy functional with energy
constraints. The mean-field entropy for the macrosate is justified
as being the opposite of a large-deviation rate function, where $N^{2}$
is the large-deviation rate. We explain those theoretical results
and their justification at a heuristic level in this paper. 

Those theoretical results (the concentration of most microstates on
a single predicted macrostate maximizing a mean-field variational
problem) provide a very interesting case for testing the Creutz's
algorithm with numerical results. This test possibility was our main
motivation for devising this algorithm for the 2D Euler equations,
before generalizing to more complex models for which mean-field type
large-deviation results are not yet available. 

We have checked that numerical results are in full agreement with
the theoretical predictions from the mean-field variational problem.
Independently of the mean-field variational problem, Creutz's algorithm
provides a very simple and robust way to sample microcanonical measures.
For instance, we have used it to describe previously unknown first-order
phase transitions between dipole and parallel flows in a doubly periodic
domain.\\

Previous works considered Monte-Carlo simulations for the 2D Euler
equations or related models (see for instance a very interesting application
to oceans in Ref. \protect\cite{Salmon2010} and references therein, or Ref.
\protect\cite{Lim}). However, those works always sampled the canonical Energy-Enstrophy
measures (using the Metropolis-Hasting algorithm and not considering
other invariants). Those method thus provide only a very small subset
of the equilibrium measures of those models. 

Dubinkina and Frank \protect\cite{Dubinkina_Frank_2010JCoPh} recently proposed
a very nice particle-mesh algorithm for the 2D Euler equations that
conserves the vorticity distribution. This algorithm is also a way
to sample microcanonical measures, as was shown in their paper. As
a positive point, the dynamics of the particle-mesh method is a good
approximation of the 2D Euler dynamics for finite times (whereas the
Creutz algorithm is just a sampling of the microcanonical measure).
One drawback of the particle-mesh algorithm is that ergodicity has
to be assumed. Moreover, it seems that this particle-mesh approach
with potential vorticity conservation has so far not been proven to
be generalizable to more complex models (for instance the Shallow
Water model).

We also note that other numerical algorithms exist to compute equilibrium
states for the 2D Euler equations: the Turkington and Whitaker algorithm
\protect\cite{turkington1,turkington2}, relaxation equations \protect\cite{robert_sommeria},
or continuation algorithms \protect\cite{bouchet_simonnet1,thess_sommeria}
(please see Ref. \protect\cite{bouchet_venaille2} for a description of those
algorithms). However, these three algorithm compute extrema or critical
points of the mean-field variational problem. They thus rely on the
large-deviation theoretical results that are so far not known to exist
for more complex models, like for instance the 3D axisymmetric equations
or the Shallow Water models.\\

In this paper we also discuss the discovery of a first-order phase
transition in the microcanonical ensemble, found using Creutz's algorithm for the 2D Euler equations.  This phase transition is striking in many respects. It is a first-order
phase transition in the microcanonical ensemble, which is a thermodynamical
curiosity (see Ref. \protect\cite{Bouchet_Barre2005}). As discussed in detail in  \protect\cite{Bouchet_Barre2005}, such a first-order phase transition in the microcanonical ensemble is a sign of ensemble inequivalence, as the entropy curve can not be concave at such a transition point. Moreover, it is a
transition from a disordered state towards an ordered state when
energy is increased, in contrast with what could be expected from
classical statistical mechanics arguments. This paradox is due to the negative temperature of the system. Indeed, then entropy, the measure of disorder, decreases when energy increases. We discuss this point further in section \protect\ref{sec:Phase-transitions-and}. From a fluid mechanics
point of view, this transition is a drastic change of the flow topology
from a dipole to a parallel flow. A very interesting recent example of two different phase transitions in two different statistical ensembles is discussed in Ref. \protect\cite{Cohen_Mukamel_2012}. \\

The outline of this paper is as follows. In Section \protect\ref{sec:Statistical-mechanics-of}
the 2D Euler equations are introduced. The statistical mechanics theory
is treated, as well as the finite-dimensional approximation of the
2D Euler microcanonical measure. Section \protect\ref{sec:Creutz-algorithm}
provides a proof of why Creutz's algorithm samples microcanonical
measures. Theoretical predictions from the microcanonical mean-field
variational problem presented in Section \protect\ref{sec:Statistical-mechanics-of}
are confronted with numerical results in Section \protect\ref{sec:Num}, where
we focus mainly on the negative temperature of the system. In Section
\protect\ref{sec:Phase-transitions-and} examples of phase transitions and
an example of ensemble inequivalence are discussed. In this section,
we also discuss the transition from a disordered state to an ordered
one upon increasing energy. Section \protect\ref{sec:Conclusion} provides
some perspectives and we give comments for future work.

\section{\label{sec:Statistical-mechanics-of}Statistical mechanics of the
2D Euler equations}

\subsection{The 2D Euler equations and invariants\label{sub:The-2D-Euler}}

The 2D Euler equations are given by 
\begin{equation}
\partial_{t}\omega+\mathbf{v}\cdot\mathbf{\nabla}\omega=0,\,\,\,\,\mathbf{v}=\mathbf{e}_{z}\times\mathbf{\nabla}\psi,\,\,\,\:\text{and}\,\,\omega=\Delta\psi,\label{eq:2deuler}
\end{equation}
where $\omega=(\mathbf{\nabla}\times\mathbf{v})\cdot\mathbf{e}_{z}$
is the vorticity, and $\mathbf{v}$ is the non-divergent velocity
expressed as the curl of the stream function $\psi$. The stream function
is defined up to a constant, which is set to zero without loss of
generality. The relation $\omega=\Delta\psi$ is complemented with
doubly-periodic boundary conditions on a domain given by $\mathscr{D}$
= $[0,1)\times[0,1)$ and $\mathbf{r}=(x,y)$. The energy of the flow
reads 
\begin{equation}
\mathscr{E}[\omega]=\frac{1}{2}\int_{\mathscr{D}}d^{2}\mathbf{r}\,\mathbf{v}^{2}=\frac{1}{2}\int_{\mathscr{D}}d^{2}\mathbf{r}\,(\nabla\psi)^{2}=-\frac{1}{2}\int_{\mathscr{D}}d^{2}\mathbf{r}\,\omega\psi.\label{eq:energy}
\end{equation}
The energy is conserved, i.e. $d_{t}\mathscr{E}=0$. The equations
also conserve an infinite number of functionals, named Casimirs. These
are related to the degenerate structure of the infinite-dimensional
Hamiltonian system and can be understood as invariants arising from
Noether's theorem \protect\cite{esm5}. These functionals are of the form
\begin{equation}
\mathscr{C}_{s}[\omega]=\int_{\mathscr{D}}d^{2}\mathbf{r}\, s(\omega),\label{eq:casimir1}
\end{equation}
where $s$ is any sufficiently smooth function depending on the vorticity
field. We note that on a doubly-periodic domain the total circulation
is zero: 

\begin{equation}
\Gamma=\int_{\mathscr{D}}d^{2}\mathbf{r}\,\omega=0.\label{eq:bd}
\end{equation}
A special Casimir is 
\begin{equation}
C(\sigma)=\int_{\mathcal{D}}d^{2}\mathbf{r}\, H(-\omega+\sigma),\label{eq:casimir2}
\end{equation}
where $H(\cdot)$ is the Heaviside step function. This Casimir returns
the area of $B(\sigma)=\left\{ \mathbf{r}\,|\,\omega(\mathbf{r})\leq\sigma\right\} ,$
i.e. the domain of all vorticity levels smaller or equal to $\sigma$.
$C(\sigma)$ is an invariant for any $\sigma$ and therefore any derivative
of it as well. Therefore, the distribution of vorticity, defined as
$D(\sigma)=C'(\sigma)$, where the prime denotes a derivation with
respect to $\sigma$, is also conserved by the dynamics. The expression
$D(\sigma)d\sigma$ is the area occupied by the vorticity levels in
the range $\sigma\leq\omega\leq\sigma+d\sigma$. 

Moreover, any Casimir can be written in the form 
\begin{equation}
\mathcal{C}_{f}[\omega]=\int d\sigma\, f(\sigma)\, D(\sigma).
\end{equation}
 The conservation of all Casimirs (Eq. (\protect\ref{eq:casimir1})) is therefore
equivalent to the conservation of $D(\sigma)$.

The conservation of the distribution of vorticity levels can also
be understood from the equations of motion, c.f. Eq. (\protect\ref{eq:2deuler}).
We find that $D\omega/dt=0$, showing that the values of the vorticity
field are Lagrangian tracers. This means that the values of $\omega$
are transported through the non-divergent velocity field, thus keeping
the distribution unchanged. The Casimirs and energy are the invariants
of the 2D Euler equations. Their existence plays a crucial role in
the dynamics of the system.

From now on, we restrict ourselves to a $K$-level vorticity distribution.
We make this choice for pedagogical reasons, but the generalization
to a continuous vorticity distribution is straightforward. The $K$-level
vorticity distribution is defined as 

\begin{equation}
D(\sigma)=\sum_{k=1}^{K}A_{k}\delta(\sigma-\sigma_{k}),\label{eq:vortdistr}
\end{equation}
where $A_{k}$ denotes the area occupied by the vorticity value $\sigma_{k}$.
The areas $A_{k}$ are not arbitrary as their sum must obey $\sum_{k=1}^{K}A_{k}=|\mathscr{D}|=1$
(the total area of the domain is equal to unity). Moreover, the boundary
condition (Eq. (\protect\ref{eq:bd})) imposes the constraint $\sum_{k=1}^{K}A_{k}\sigma_{k}=0$.

\subsection{Microcanonical measure \label{sub:Microcanonical-measure-of}}

As the 2D Euler equation is a conservation law (the time derivative of $\omega$ is the divergence of a current), it verifies a formal Liouville theorem \protect\cite{bouchet_thalabard,robert3}. This formally justifies that the microcanonical measure will be dynamically invariant. In order to properly identify a microcanonical measure, we will thus discretize the vorticity field in
finite-dimensional space, with  $N^2$ degrees of freedom, and then take the limit $N\rightarrow\infty$. As any point in the physical space has a symmetric role, a uniform grid has to be chosen in order to preserve the volume conservation corresponding to the Liouville theorem.

We denote the lattice points by $\mathbf{r}_{ij}=\left(\frac{i}{N},\frac{j}{N}\right)$,
with $0\leq i,j\leq N-1$ and denote $\omega_{ij}^{N}=\omega(\mathbf{r}_{ij})$
to be the vorticity value at point $\mathbf{r}_{ij}$. The total number
of points is $N^{2}$. 

As discussed in the previous section, we assume $D(\sigma)=\sum_{k=1}^{K}A_{k}\delta(\sigma-\sigma_{k})$.
For this finite-$N$ approximation, our set of microstates (configuration
space) is then

\begin{equation}
X_{N}=\left\{ \omega^{N}=(\omega_{ij}^{N})_{0\leq i,j\leq N-1}\mid\forall i,j\;\omega_{ij}^{N}\in\left\{ \sigma_{1},\ldots,\sigma_{K}\right\} ,\;\mbox{and}\,\,\forall k\,\,\,\#\left\{ \omega_{ij}^{N}\mid\omega_{ij}^{N}=\sigma_{k}\right\} =N^{2}A_{k}\right\} .\label{eq:config}
\end{equation}
Here, $\#(A)$ is the cardinality of set $A$. We note that $X_{N}$
depends on $D(\sigma)$ through $A_{k}$ and $\sigma_{k}$ (see Eq.
(\protect\ref{eq:vortdistr})). We note that all microstates in $X_{N}$ have
the proper vorticity distribution. 

In order to define a microcanonical ensemble, we also have to impose
an energy constraint. Using the above expression we define the energy
shell $\Gamma_{N}(E,\Delta E)$ as

\begin{equation}
\Gamma_{N}(E,\Delta E)=\left\{ \omega^{N}\in\Gamma_{N}\mid E_{0}\leq\mathscr{E}_{N}[\omega^{N}]\leq E_{0}+\Delta E\right\} ,\label{eq:microensemble}
\end{equation}
where 

\begin{equation}
\mathscr{E}_{N}=\frac{1}{2N^{2}}\sum_{i,j=0}^{N-1}\left(\mathbf{v}_{ij}^{N}\right)^{2}=-\frac{1}{2N^{2}}\sum_{i,j,i',j'=0}^{N-1}\omega_{ij}^{N}G_{ij,i'j'}\omega_{i'j'}^{N}\label{eq:energy-1}
\end{equation}
is the finite-$N$ approximation of the system energy, with $\mathbf{v}_{ij}^{N}=\mathbf{v}(\mathbf{r}_{ij})$
being the discretized velocity field, $\Delta E$ is the width of
the energy shell, and where $G_{ij,i'j'}$ is a finite$-N$ approximation
of the Laplacian Green function on domain $\mathscr{D}$. We shall
define these finite-$N$ approximate fields more precisely in Section
\protect\ref{sec:Creutz-algorithm}. Note that a finite width of the energy
shell is necessary for our discrete approximation, as the cardinality
of $X_{N}$ is finite. Indeed, the set of accessible energies on $X_{N}$
is also finite. Let $\Delta_{N}E$ be the typical difference of between
two successive achievable energies. We therefore assume that $\Delta_{N}E\ll\Delta E\ll E_{0}$. 

The fundamental assumption of statistical mechanics states that all
microstates in this ensemble are equiprobable. By virtue of this assumption,
the probability to observe any microstate is $\Omega_{N}^{-1}(E_{0},\Delta E)$,
where $\Omega_{N}(E_{0},\Delta E)$ is the number of accessible microstate
and is defined as the cardinality of the set $\Gamma_{N}(E_{0},\Delta E)$.
The finite-$N$ specific Boltzmann entropy is then given by
\begin{equation}
S_{N}(E_{0},\Delta E)=\frac{1}{N^{2}}\log\Omega_{N}(E_{0},\Delta E).\label{eq:specificboltzmann}
\end{equation}
The microcanonical measure is defined through the expectation values
of any observable $\mathscr{A}$. For any observable $\mathscr{A}[\omega]$
(for instance a smooth functional of the vorticity field), we refer
to its finite-$N$ approximation by $\mathscr{A}_{N}[\omega^{N}]$.
The expectation value of $\mathscr{A}_{N}$ for the microcanonical
measure reads 

\begin{equation}
\left\langle \mu_{N}(E_{0},\Delta E),\,\mathscr{A}_{N}(\omega^{N})\right\rangle \equiv\left\langle \mathscr{A}_{N}(\omega^{N})\right\rangle _{N}\mbox{\ensuremath{\equiv}}\frac{1}{\Omega_{N}(E_{0},\Delta E)}\sum_{\omega^{N}\in\Gamma_{N}(E_{0},\Delta E)}\mathscr{A}_{N}(\omega^{N}).
\end{equation}
The microcanonical measure $\mu$ for the 2D Euler equation is defined
as a limit of the finite-$N$ measure: 

\begin{equation}
<\mu(E_{0}),\,\mathscr{A}(\omega)>\,\equiv\lim_{N\rightarrow\infty}<\mu_{N}(E_{0},\Delta E),\,\mathscr{A}_{N}(\omega^{N})>.
\end{equation}
The specific Boltzmann entropy is defined as

\begin{equation}
S(E_{0})=\lim_{N\rightarrow\infty}S_{N}(E_{0},\Delta E).\label{eq:boltzmann}
\end{equation}
The limit measure and the entropy just defined are expected to be
independent of $\Delta E$ in the limit $N\rightarrow\infty$. This
will be justified in next section with large-deviation principles.

\subsection{Sanov theorem and mean-field entropy\label{sub:Sanov-theorem}}

Computing the Boltzmann entropy by direct evaluation of Eq. (\protect\ref{eq:boltzmann})
is usually an intractable problem. However, we shall give heuristic arguments in order to show that the limit $N\rightarrow\infty$ can be easily evaluated.  The Boltzmann entropy (Eq. (\protect\ref{eq:boltzmann}))
can then be computed through maximizing a constraint variational problem
(called a mean-field variational problem, see Eq. (\protect\ref{eq:var})). 

This variational problem is the foundation of the Robert-Sommeria-Miller
(RSM) approach to the equilibrium statistical mechanics for the 2D
Euler equations. The essential message is that the entropy computed
from the mean-field variational problem (to be defined below) and
from Boltzmann's entropy definition (Eq. (\protect\ref{eq:boltzmann})) are
equal in the limit $N\rightarrow\infty$. The ability to compute the
Boltzmann entropy through this type of variational problems is one
of the cornerstones of statistical mechanics.

Our heuristic derivation is based on the same type of combinatoric
argument as the ones used by Boltzmann for the interpretation of his
$H$ function in the theory of relaxation to equilibrium of a dilute
gas. This derivation doesn't use the technicalities of large-deviation
theory. The aim is to actually obtain the large-deviation interpretation
of the entropy and to provide a heuristic understanding using basic
mathematics only. The modern mathematical proof of the relation between
the Boltzmann entropy and the mean-field variational problem involves
the theory of large deviations and Sanov's theorem. 

Macrostates are sets of microscopic configurations sharing similar
macroscopic behavior. Our aim is to properly identify macrostates
that fully describe the main features of the largest scales of 2D
turbulent flows and computing their probability or entropy.\linebreak{}

Let us first define macrostates through local coarse-graining. We
divide the $N\times N$ lattice into $(N/n)\times(N/n)$ non-overlapping
boxes each containing $n^{2}$ grid points ($n$ is an even number,
and $N$ is a multiple of $n$). These boxes are centered on sites
$(i,j)=(In,Jn),$ where integers $I$ and $J$ verify $0\leq I,J\leq N/n-1$.
The indices $(I,J)$ label the boxes. 

For any microstate $\omega^{N}\in\Gamma_{N}$, let $F_{k,IJ}^{N}$
be the frequency to find a vorticity value $\sigma_{k}$ in box $(I,J)$:
\[
\label{eq:F-k}
F_{k,IJ}^{N}(\omega^{N})=\frac{1}{n^{2}}\sum_{i=In-n/2+1}^{In+n/2}\,\sum_{j=Jn-n/2+1}^{Jn+n/2}\delta_{d}(\omega_{ij}^{N}-\sigma_{k}),
\]
where $\delta_{d}(x)$ is equal to one whenever $x=0$, and zero otherwise.
We note that for all $(I,J)$ $\sum_{k}F_{k,IJ}^{N}(\omega^{N})=1$. 

A macrostate $p^{N}=\left\{ p_{k,IJ}^{N}\right\} _{0\leq I,J\leq N/n-1;1\leq k\leq K}$,
is the set of all microstates $\omega^{N}\in X_{N}$ such that $F_{k,IJ}^{N}(\omega^{N})=p_{k,IJ}^{N}$
for all $I,J$, and $k$ (by abuse of notation, and for simplicity,
$p^{N}=\left\{ p_{k,IJ}^{N}\right\} _{0\leq I,J\leq N/n-1;1\leq k\leq K}$
refers to both the set of values $p_{k,IJ}^{N}$ and to the set of
microstates having the corresponding frequencies). The property $\sum_{k}F_{k,IJ}^{N}(\omega^{N})=1$ imposes a local
normalization constraint $\forall I,J\;\sum_{k}p_{k,IJ}^{N}=1$.
The entropy of the macrostate is defined as the logarithm of the number
of microstates in the macrostate 
\begin{equation}
S_{N}[p^{N}]=\frac{1}{N^{2}}\log\#\left\{ \omega^{N}\in X_{N}\left|\,\,\,\mbox{for\,\,\ all\,\,}I,J,\,\mbox{and}\,\,\, k,\,\,\, F_{k,IJ}(\omega^{N})=p_{k,IJ}^{N}\right.\right\} .\label{eq:entropy4}
\end{equation}
From an argument by Boltzmann (a classical exercise in statistical
mechanics), using combinatorics and the Stirling formula, the limit
$N>n\gg1$, the asymptotical entropy of the macrostate is
\[
S_{N}[p^{N}]\underset{N\gg n\gg1}{\sim}\begin{cases}
\mathscr{S}_{N}[p^{N}]=-\frac{n^{2}}{N^{2}}\sum_{I,J=0}^{N/n-1}\sum_{k=1}^{K}p_{k,IJ}^{N}\log p_{k,IJ}^{N} & \mbox{if}\,\,\,\forall I,J\;\mathscr{N}[p_{IJ}]=1\,\text{and}\,\forall k\,\mathscr{A}_{N}[p_{k}^{N}]=A\\
-\infty & \mbox{otherwise},
\end{cases}
\]
where $\mathscr{N}[p_{IJ}^{N}]\equiv\sum_{k}p_{k,IJ}^{N}$. In large-deviation
theory, this result could have been obtained using Sanov's theorem. 

The coarse-grained vorticity is defined as 
\begin{equation}
\overline{\omega_{IJ}^{N}}=\frac{1}{n^{2}}\sum_{i'=In-n/2+1}^{In+n/2}\sum_{j'=Jn-n/2+1}^{Jn+n/2}\omega_{i'j'}^{N}.\label{eq:cgomega}
\end{equation}
Note that, over the macrostate $p^{N}$, the coarse-grained vorticity
depends on $p^{N}$ only:
\[
\overline{\omega_{IJ}^{N}}=\sum_{k=1}^{K}p_{k,IJ}^{N}\sigma_{k}\,\,\,\mbox{for\,\,\,}\omega^{N}\in p^{N}.
\]
We now consider a new macrostate $(p^{N},E_{0})$ which is the set
of microstates $\omega^{N}$ with energy $\mathscr{E}_{N}[\omega^{N}]$
verifying $E_{0}\leq\mathscr{E}_{N}[\omega^{N}]\leq E_{0}+\Delta E$
(the intersection of $\Gamma_{N}(E,\Delta E)$ and $p^{N}$). For
a given macrostate $p^{N}$, not all microstates have the same energy.
Thus, the constraint on the microstate energy cannot be recast as
a simple constraint on the macrosate $p^{N}$. Therefore, treating
the energy constraint requires a more subtle approach. The energy
(\protect\ref{eq:energy-1}) is
\begin{equation}
\mathscr{E}_{N}[\omega^{N}]=-\frac{1}{2N^{4}}\sum_{i,j,i',j'=0}^{N-1}\omega_{ij}^{N}G_{ij,i'j'}\omega_{i'j'}^{N}.\label{eq:energy_green}
\end{equation}
Then, in the limit $N\gg n\gg1$, the variations of $G_{ij,i'j'}$
for $(i',j')$ running over the small box $(I,J)$ are vanishingly
small. Hence, $G_{ij,i'j'}$ can be well approximated by the average
value over the boxes $G_{IJ,I'J'}$

\begin{equation}
G_{ij,i'j'}=G_{IJ,I'J'}+o\left(\frac{1}{n}\right).\label{eq:green_laplace}
\end{equation}
From Eq. (\protect\ref{eq:energy_green}), using Eqs. (\protect\ref{eq:cgomega},
\protect\ref{eq:green_laplace}), it is easy to conclude that in the limit
$N\gg n\gg1$ the energy of any microsate of the macrosate $p^{N}$
is very well approximated by the energy of the coarse-grained vorticity
\[
\mathscr{E}_{N}[\omega^{N}]\underset{N\gg n\gg1}{\sim}\mathscr{E}_{N}[\overline{\omega_{IJ}^{N}}]=-\frac{n^{2}}{2N^{2}}\sum_{I,J=0}^{N/n-1}\overline{\omega_{IJ}^{N}}\psi_{IJ}^{N},
\]
where $\psi_{IJ}^{N}$ is the stream function, and is related to the
velocity field and vorticity field by Eqs. (\protect\ref{eq:2deuler}). Note
that in the above equation we made use of the relation $\mathscr{E}[\omega^{N}]=\mathscr{E}_{N}[\overline{\omega^{N}}]+o\left(\frac{1}{n}\right)$. 

Hence, the Boltzmann entropy of the macrostate is
\begin{equation}
S_{N}[p^{N},E_{0}]\underset{N\gg n\gg1}{\sim}\begin{cases}
\mathscr{\mathscr{S}}_{N}[p^{N}] & \mbox{if}\,\,\,\forall k\;\mathscr{A}_{N}\left(p_{k}^{N}\right)=A_{k},\,\mathscr{N}[p^{N}]=1\,\mbox{and}\,\,\,\mathscr{E}_{N}[\overline{\omega_{IJ}^{N}}]=E_{0}\\
-\infty & \mbox{otherwise.}
\end{cases}\label{eq:specialdeviation}
\end{equation}
Consider $P_{N,E_{0}}(p^{N})$ to be the probability density to observe
the macrostate $p^{N}$. By definition of the microcanonical ensemble
and of the entropies $S_{N}(E_{0})$ (see Eq. (\protect\ref{eq:specificboltzmann}))
and $S_{N}(p^{N},E_{0})$ (see Eq. (\protect\ref{eq:specialdeviation})),
we have

\begin{equation}
P_{N,E_{0}}(p^{N})=\exp\left\{ N^{2}\left[S_{N}[p^{N},E_{0}]-S_{N}(E_{0})\right]\right\} .\label{eq:largedeviationresult}
\end{equation}

Let $P_{M}$ be the probability density for the random variable $X_{M}$.
The statement 

\begin{equation}
\lim_{M\rightarrow\infty}-\frac{1}{M}\log\left[P_{M}(X_{M}=x)\right]=I(x)
\end{equation}
is called a large-deviation result. $I(x)$ is the large-deviation
rate function, and $M$ the large-deviation rate. From this definition,
we see that formula (\protect\ref{eq:largedeviationresult}) is a large-deviation
result for macrostate $p^{N}$ for the macrocanonical measure. The
large-deviation rate is $N^{2}$ and the large-deviation rate function
is $-S_{N}[p^{N},E_{0}]+S_{N}(E_{0})$.\\

We now consider the continuous limit $n\rightarrow\infty,\, N\rightarrow\infty$.
The macrostates $p_{k}^{N}$ are now seen as finite-$N$ approximation
of $p_{k}$, the local probability to observe $\omega(\mathbf{r})=\sigma_{k}:\, p_{k}(\mathbf{r})=\left\langle \delta(\omega(\mathbf{r})-\sigma_{k})\right\rangle $.
The macrostate is now characterized by $p=\left\{ p_{1},\ldots,p_{K}\right\} $.
Taking the limit $N\gg n\gg1$ allows us to define the entropy of
the macrostate $(p,E_{0})$ as
\begin{equation}
S[p,E_{0}]\underset{}{=}\begin{cases}
\mathscr{\mathscr{S}}[p]\equiv & -\sum_{k}\int_{\mathscr{D}}d\mathbf{r}\, p_{k}\mbox{\ensuremath{\log}}p_{k}\,\,\mbox{if}\,\,\,\forall k\;\mathscr{N}[p^{k}]=1,\,\,\,\mathscr{A}\left(p^{k}\right)=A_{k}\,\,\,\mbox{and}\,\,\,\mathscr{E}[\overline{\omega}]=E_{0}\\
-\infty & \mbox{otherwise,}
\end{cases}\label{eq:entropy5}
\end{equation}
where $\forall\mathbf{r}$, $\mathscr{N}(\mathbf{r})=\sum_{k=1}^{K}p_{k}(\mathbf{r)}=1$
is the local normalization. In the same limit, it is clearly seen
from definition (\protect\ref{eq:entropy4}) and result (\protect\ref{eq:entropy5})
that there is a concentration of microstates close to the most probable
macrostate: the equilibrium state. The exponential concentration close
to the equilibrium state is a large-deviation result, where the entropy
appears as the opposite of a large-deviation rate function (up to
a constant). 

The exponential convergence towards this most probable state also
justifies the approximation of the above entropy with the entropy
of the most probable macrostate, Eq. (\protect\ref{eq:boltzmann}), as

\begin{equation}
S(E_{0})=\max_{\left\{ p\right\} \mid\mathscr{N}(\mathbf{r})=1}\left\{ \mathscr{\mathscr{S}}[p]\,\mid\mathscr{E}[\bar{\omega}]=E_{0},\,\forall k\;\mathscr{A}(p^{k})=A_{k}\right\} ,\label{eq:var}
\end{equation}
where $p=\left\{ p_{1},\ldots,p_{K}\right\} $, $\forall\mathbf{r}$
$\mathscr{N}(\mathbf{r})=\sum_{k=1}^{K}p_{k}(\mathbf{r)}=1$ is the
local normalization, $\mathscr{S}[p]$ is as defined in Eq. (\protect\ref{eq:entropy5}),
and $\mathscr{A}[p_{k}]$ is the area of the domain corresponding
to the vorticity value $\omega=\sigma_{k}$. The fact that the Boltzmann
entropy $S(E_{0})$ (Eq. (\protect\ref{eq:boltzmann})) can be computed from
the variational problem (\protect\ref{eq:var}) is a powerful non-trivial
result from large-deviation theory.

In the next section, we shall test the prediction of concentration
of microstates close to the equilibrium macrostate with numerical
simulations. We first define Creutz's algorithm and explain why it
is able to sample microcanonical measures. We continue by applying
this algorithm to the 2D Euler equations.

\section{\label{sec:Creutz-algorithm}Creutz's algorithm}

Creutz's algorithm was introduced by M. Creutz in 1983 \protect\cite{creutz}.
It is a generalization of the Metropolis-Hastings algorithm that samples
the Gibbs measure (with a Boltzmann factor). Creutz's algorithm, on
the other hand, samples the microcanonical measure in the energy shell
$E_{0}\leq\mathscr{E}\leq E_{0}+\Delta E$ (a uniform distribution
over the set $\Gamma_{N}(E_{0},\Delta E)$). Here, the system energy
is denoted by $\mathscr{E}$. 

In Section \protect\ref{sub:Definition-of-the} we present Creutz's algorithm
and prove that it actually samples the microcanonical measure. In
Section \protect\ref{sub:Temperature-computation-from} we provide a method
to calculate the inverse temperature $\beta$ using this algorithm.
We shall refer to Appendix A for a precise definition of Markov chains,
the detailed balance condition, and invariant measures (stationary
distributions). \\

$\textbf{Creutz and others using his algorithm used
the notion of a daemon.}$ The aim of the daemon is to allow
for slight energy fluctuations, which are necessary for systems with
a discrete configuration space. With our notation the daemon energy
$E_{d}$ is nothing else than $E_{d}=E_{0}+\Delta E-\mathcal{E}$.
The original Creutz algorithm samples a uniform measure over all microstates
of energy smaller than $E_{0}+\Delta E$. For this measure, in systems
with positive temperature and a large number of degrees of freedom,
the energy distribution is concentrated close to $E_{0}+\Delta E$,
and typical energy fluctuations are small. 

In the 2D Euler case considered in this paper the temperature can
be negative, as will be discuss in Section \protect\ref{sub:Negative-temperature-in}.
Microstates with the smallest possible energy then become overwhelmingly
probable. In order to sample the microcanonical measure, we then need
to impose a lower-bound on the energy. In order to cope with all possible
temperature cases we sample a uniform measure over the energy shell
$E_{0}\leq\mathscr{E}\leq E_{0}+\Delta E$. Because of energy concentration
properties, the microcanonical measure is independent on those definitions
or on the value of $\Delta E$ in the limit of an infinite number
of degrees of freedom.

Classic heuristic arguments using the daemon energy lead to misleading
conclusions, for instance in the case of negative-temperature systems.
For this reason we prefer not to use this concept at all. We notice
moreover that the deamon concept is actually not useful as its energy
contains no more information than the state energy $\mathcal{E}$.

\subsection{Definition of Creutz's algorithm\label{sub:Definition-of-the}}

We consider an ensemble of states of a physical system. Each state
$x$ is a set of $M$ values: $x=(x_{i})_{1\leq i\leq M}$. The set
of states

\begin{equation}
X=\left\{ x=(x_{i})_{1\leq i\leq M}\right\} 
\end{equation}
is called the $\textit{configuration space}$. Note that $i\in\mathbb{N}$
is the index for components of $x$.

For instance, a state of an Ising spin system is given by $\forall i\; x_{i}\in\left\{ -\frac{1}{2},+\frac{1}{2}\right\} $.
Note that the values $x$ could also be continuous, i.e. $\forall i\; x_{i}\in\mathbb{R}$.
For the 2D Euler equations, the configuration space is $X_{N}$ as
defined in Eq. (\protect\ref{eq:config}), and $M=N^{2}$.

The energy of each microstate is given by $\mathscr{E}(x)$. Furthermore,
we define

\begin{equation}
\Gamma(E_{0},\Delta E)=\left\{ x\in X\mid E_{0}\leq\mathscr{E}(x)\leq E_{0}+\Delta E\right\} 
\end{equation}
as the set of microstates in the energy shell $E_{0}\leq\mathscr{E}(x)\leq E_{0}+\Delta E$.
The microcanonical measure is defined as the uniform measure over
$\Gamma(E_{0},\Delta E)$ (each microstate in $\Gamma(E_{0},\Delta E)$
has probability $\Omega(E,\Delta E)=(\#\Gamma(E_{0},\Delta E))^{-1}$
to occur, where we recall that $\#(A)$ is the cardinality of set $A$).
Furthermore, as in the previous section, we assume that $\Delta_{M}E\ll\Delta E\ll E_{0}$.
Our goal is to sample the microcanonical measure over $\Gamma(E_{0},\Delta E)$.

We assume that there exists a Markov chain $\mathscr{T}$, defined
through a sequence of random numbers $\left\{ y^{l}\in X\right\} _{l\geq0}$
with $\textit{transition probability}$ $T(x,x')$ (so $T(x,x')$
is the probability to observe $y^{l+1}=x$ if $y^{l}=x'$). Here,
$l\in\mathbb{N}$ is the index for the position in the Markov chain.
We assume that $\mathscr{T}$ verifies detailed balance for the uniform
measure on $X$. This is equivalent to the statement $T(x,x')=T(x',x)$.
See Appendix A for more details. Lastly, we note that $\mathscr{T}$
(or $T$) does not depend on the system energy.

In order to sample the microcanonical measure over $\Gamma(E_{0},\Delta E)$
an algorithm is needed that generates realizations $\left\{ z^{l}\right\} $
of a new Markov chain $\mathscr{Q}$ defined by its corresponding
transition probability $Q$ and over the configuration space $\Gamma(E_{0},\Delta E)$.
We shall call this algorithm Creutz's algorithm, and we define it
in the following way.

Let the system's current state be $y^{l}=x'$. We pick at random $x\in X$
with probability $T(x,x')$. If $E_{0}\leq\mathscr{E}(x)\leq E_{0}+\Delta E$
then we accept this move and $y^{l+1}=x$. Otherwise, we do not accept
and $y^{l+1}=y^{l}=x'$. Iteration of this procedure defines a Markov
chain $\mathscr{Q}$.

We can show that $Q$ verifies detailed balance for the uniform measure
over $\Gamma(E_{0},\Delta E)$. It is easily checked that $Q(x,x')=T(x,x')$
if $x\neq x'$, and $Q(x,x)=T(x,x)+\sum_{x'\in X\backslash\Gamma(E,\Delta E)}T(x',x)=1-\sum_{x'\in\Gamma(E_{0},\Delta E),x\neq x'}T(x,x')$.
($Q(x,x)$ is the probability of a move from $y^{l}$ to $y^{l+1}$
to fail because of the energy constraint.) We thus find that $\forall x,x'\in\Gamma(E_{0},\Delta E):\; Q(x,x')=Q(x',x)$
by virtue of the detailed balance on $\mathscr{T}$. Therefore, $\mathscr{Q}$
verifies a detailed balance for a uniform measure on $\Gamma(E_{0},\Delta E)$.
Thus, $\mathscr{Q}$ has this uniform measure as a stationary measure,
see appendix A. In conclusion, if $\mathscr{Q}$ is ergodic then it
samples the microcanonical measure over $\Gamma(E_{0},\Delta E)$.\\

We conclude by describing how to properly empirically sample an observable
$A$. The expectation value of observable $A$ is computed through

\begin{equation}
<A(y)>=\lim_{N\rightarrow\infty}\frac{1}{L}\sum_{l=1}^{L}A(y^{l}).\label{eq:sumobs}
\end{equation}
It is important to notice that if for $y^{l}$, the Creutz algorithm
fails to change the state $K$ times ($y^{l}=y^{l+1}=\ldots y^{l+K-1}$),
then these $K$ configurations need to be included into the sum (\protect\ref{eq:sumobs}).
We clarify this statement in Appendix B.

\subsection{Temperature computation using the Creutz algorithm\label{sub:Temperature-computation-from}}

We continue with the computation of the inverse temperature from the
energy distribution at equilibrium. We denote $\rho_{M}(E)$ as the
density of states for energy $E$. The number of microstates in $\Gamma(E_{0},\Delta E)$
then reads

\begin{equation}
\Omega_{M}(E_{0},\Delta E)=\int_{E_{0}}^{E_{0}+\Delta E}dE\,\rho_{M}(E).\label{eq:laplace_integral}
\end{equation}
We assume that the energy density of state has a large deviation behavior,
i.e.,

\begin{equation}
\forall E\,\,\,\,\lim_{M\rightarrow\infty}\frac{1}{M}\log\rho_{M}(E)\underset{M\rightarrow\infty}{\rightarrow}S(E).
\end{equation}
We can assume a stronger property, namely

\begin{equation}
\rho_{M}(E)\underset{M\rightarrow\infty}{\sim}C(E)\, e^{MS(E)}.
\end{equation}
We note that this assumption is verified for most microcanonical measures.
For the 2D Euler equations, it follows from the large-deviation results
discussed in Section \protect\ref{sub:Sanov-theorem}.

In the microcanonical ensemble the temperature is defined as $\beta=\frac{dS}{dE}$.
It then follows that $S(E)=S(E_{0})+\beta(E-E_{0})+o(E-E_{0})$ and
$\rho_{M}(E)\underset{M\rightarrow\infty}{\sim}C(E)e^{MS(E_{0})+\beta M(E-E_{0})}$.
We thus have, using the fact that Eq. (\protect\ref{eq:laplace_integral})
is a Laplace integral, that $\lim_{M\rightarrow\infty}\frac{1}{M}\log\Omega_{M}(E_{0},\Delta E)=S(E_{0})$
is independent of $\Delta E$.

Moreover, 

\begin{equation}
\rho_{M}(E)\underset{M\rightarrow\infty}{\sim}C_{1}(E_{0},\Delta E)\, e^{-\beta M(E-E_{0})},\label{eq:betadistr}
\end{equation}

and

\begin{equation}
P(E)\underset{M\rightarrow\infty}{\sim}C_{2}(E_{0},\Delta E)\, e^{-\beta M(E-E_{0})}.
\end{equation}
The energy distribution is exponential with rate $\beta M$. If $\beta$
is positive, the energy of the system is concentrated close to $E_{0}+\Delta E$.
If $\beta$ is negative, the energy of the system is concentrated
close to $E_{0}$. 

We will return to this result in Section \protect\ref{sec:Num}.

\subsection{2D Euler algorithm\label{sub:2D-Euler-algorithm}}

We now want to apply the Creutz algorithm to the 2D Euler model. We
will follow the general definition from Section \protect\ref{sub:Definition-of-the}.
Thus, we need to (i) define a Markov chain for the 2D Euler model
and (ii) precisely define the approximate energy such that we can
sample states in the energy shell $[E_{0},E_{0}+\Delta E]$. 

Firstly, following the previous section, we consider the Markov chain
$\mathscr{T}$, now defined through its configuration space

\begin{equation}
X_{N}=\left\{ \omega^{N}=(\omega_{ij})_{0\leq i,j\leq N-1}\mid\forall i,j\;\omega_{ij}^{N}\in\left\{ \sigma_{1},\ldots,\sigma_{n}\right\} ,\,\text{and}\forall k\,\#\left\{ \omega_{ij}^{N}\mid\omega_{ij}^{N}=\sigma_{k}\right\} =N^{2}A_{k}\right\} ,
\end{equation}
and transition probability $T(\omega,\omega')$. $T$ denotes the
probability to go from a vorticity state $y^{l}=\omega'\in X_{N}$
to $y^{l+1}=\omega\in X_{N}$, where as before $l\geq0$ denotes the
position of $y^{l}$ in $\mathscr{T}$. We assume the detailed balance
condition holds for $\mathscr{T}$.

Secondly, we need to calculate the approximate energy $\mathscr{E}_{N}$
given by Eq. (\protect\ref{eq:energy-1}). The discretized vorticity field
$\omega^{N}$ is transformed to Fourier space with a two-dimensional
(discrete) Fast Fourier Transform (FFT). The discretized velocity
field and stream function field are then computed by making use of
the Fourier representation of Eqs. (\protect\ref{eq:2deuler}) and an inverse
FFT, resulting in $\mathbf{v}^{N}=(\mathbf{v}_{ij}^{N})_{0\leq i,j\leq N-1}$
and $\psi^{N}=(\psi_{ij}^{N})_{0\leq i,j\leq N-1}$.

With these definitions the microcanonical measure over Eq. (\protect\ref{eq:microensemble})
can be sampled. We take the current system state to be $y^{l}=\omega\in X_{N}$
(the current vorticity field configuration) and $y^{l+1}=\omega\in X_{N}$
as the next state (next vorticity field configuration) in $\mathscr{T}$.
Similarly, the current energy is denoted by $\mathscr{E}_{N}[y^{l}]$
and the energy of the next state by $\mathscr{E}_{N}[y^{l+1}]$. 

We now describe how we construct a realization of the Markov chain
$\mathscr{T}$ under the energy constraint. The vorticity field $y^{0}=\omega^{N}\in X_{N}$
is initialized with $N^{2}/2$ values of $\sigma_{+}=+1$ and $N^{2}/2$
values of $\sigma_{-}=-1$ such that conditions $\sum_{k=1}^{2}A_{k}=1$
and $\sum_{i,j=0}^{N-1}\omega_{ij}^{N}=0$ are satisfied (see Section
\protect\ref{sub:The-2D-Euler}). We use a two-level potential vorticity distribution
as an example, but generalization to higher-level distributions is
straightforward. To sample the microcanonical measure, obeying the
energy and vorticity distribution constraints, the following routine
is followed:
\begin{enumerate}
\item Let the current state of the system be $y^{l}=\omega'$. We randomly
chose two lattice sites, $(i_{0},j_{0})$ and $(i_{1},j_{1})$ with
a uniform distribution. The values at these sites are $\omega'_{i_{0}j_{0}}$
and $\omega'_{i_{1}j_{1}}$, respectively. The vorticity values at
these positions are then interchanged. In other words: $\omega_{ij}=\omega'_{ij}$
$\forall ij\neq\left\{ i_{0}j_{0},\, i_{1}j_{1}\right\} $, $\omega_{i_{0}j_{0}}=\omega'_{i_{1}j_{1}}$,
and $\omega{}_{i_{1}j_{1}}=\omega'_{i_{0}j_{0}}$. We note that by
interchanging two vorticity values, the new field $\omega$ still
belongs to $X_{N}$.
\item The new energy, $\mathscr{E}_{N}[\omega^{N}]$, is calculated. 
\item The energy check is performed. If $E_{0}\leq\mathscr{E}_{N}[\omega^{N}]\leq E_{0}+\Delta E$
then we accept the move in step 1, i.e., $\omega^{l+1}=\omega$. If
the new energy is not in the allowed energy range, we reverse step
1 such that we get the old vorticity field back, i.e., $\omega^{l+1}=\omega^{l}=\omega$'.
The new energy in this case is thus given by $\mathscr{E}_{N}[\omega]=\mathscr{E}_{N}[\omega']$.
This step ensures conservation of energy and only configurations are
allowed which are in the set $\Gamma_{N}(E_{0},\Delta E)$, see Eq.
(\protect\ref{eq:microensemble}). In either case, we return to step 1.
\end{enumerate}
Iteration of steps (1-3) then builds a realization $\left\{ \omega^{l}\right\} _{l\geq0}$
of the Markov chain $Q$. Since we assumed that detailed balance holds
for $\mathscr{T}$, we know (by virtue of Section \protect\ref{sub:Definition-of-the})
that the microcanonical measure is an invariant measure of $Q$. If
we assume $Q$ to be ergodic then we sample the microcanonical measure.

This vorticity exchange method has some analogies with spin ones in the Kawasaki algorithm \protect\cite{Newman_Barkema_1999}. The Kawasaki algorithm is a convenient way to treat dynamics with conservation laws in spin or lattice gas Monte Carlo dynamics. Our algorithm slightly differs from Kawasaki's, as we consider non-local vorticity value exchanges.

We define a Monte-Carlo step as $N^{2}$ accepted changes in the vorticity
field, for reasons mentioned in Section \protect\ref{sub:Definition-of-the}.
Typically, equilibrium is reached after ten Monte-Carlo steps for
a two-level distribution and fifty for a three-level vorticity distribution.
We note that the equilibration time depends on the energy $E_{0}$
chosen.

\section{Numerical results\label{sec:Num}}

In this section we present the numerical results obtained by sampling
the microcanonical measure of the 2D Euler equations using the algorithm
described in the previous section. For pedagogical reasons, we restrict
the simulations to two- and three-level vorticity distributions, but
note that generalization to more general vorticity level distributions
is straightforward. We consider the vorticity distribution (see Eq.
(\protect\ref{eq:vortdistr})) 
\begin{equation}
D(\sigma)=\frac{1}{2}\,\delta(\sigma-\sigma_{+})+\frac{1}{2}\,\delta(\sigma-\sigma_{-})
\end{equation}
for the two-level distribution, and 

\begin{equation}
D(\sigma)=\frac{1}{4}\delta(\sigma-\sigma_{+})+\frac{1}{2}\delta(\sigma-\sigma_{0})+\frac{1}{4}\delta(\sigma-\sigma_{-})
\end{equation}
for the three-level distribution. In the above, $\sigma_{+}=1,\,\sigma_{0}=0,\,\sigma_{-}=-1$.

\subsection{Mean-field behavior and negative temperature}

\subsubsection{Mean-field predictions\label{sub:Mean-field-predictions}}

The variational problems (\protect\ref{eq:var}) associated to the two- and
three-level cases have been explicitly solved in the past, see Refs.
\protect\cite{robert_sommeria,thess_sommeria,juttner_thess}. We summarize
their results here and compare them to our numerical results. 

For the two-level vorticity case the local probability to find a vorticity
value of $\omega=\sigma_{+}=1$ is found to be $\rho_{1}=\frac{1}{2}(1-\tanh(\alpha-\beta\psi(\mathbf{r})))$
\protect\cite{bouchet_sommeria1}, where $\alpha$ and $\beta$ are Lagrange
parameters associated with the conservation of area $A$ and energy,
respectively. Furthermore, $\beta$ is the inverse temperature: $\beta=\frac{dS}{dE}$.
Since there are only two possible values for $\omega$, the local
probability to find a vorticity value of $\omega=\sigma_{-}=-1$ is
therefore $\rho_{2}=1-\rho_{1}$. Hence, the locally averaged vorticity
reads

\begin{equation}
\overline{\omega}=\rho_{1}\sigma_{+}+(1-\rho_{1})\sigma_{-}=-\tanh(\alpha-\beta\psi).\label{eq:fit2v}
\end{equation}
We consider a symmetric distribution $D(\sigma)$ ($D(\sigma)=-D(-\sigma)$).
If we assume this symmetry then $\alpha=0$. 

For the three-level distribution a similar calculation, assuming symmetry
is not broken, yields \protect\cite{juttner_thess}

\begin{equation}
\overline{\omega}=\frac{\mu\sinh(\beta\psi)}{1+\mu\cosh(\beta\psi)},\label{eq:fit3v}
\end{equation}
where $\mu$ is an additional Lagrange parameters arising from the
conservation of $A$.

\subsubsection{Negative temperature in doubly periodic domains\label{sub:Negative-temperature-in}}

The first statistical mechanics approach to self organization of 2D
turbulence was Onsager's work on the point vortex model \protect\cite{onsager1949}.
Onsager argued in his paper that ensembles of point vortices may have
a negative temperature. This property is traced back to the fact that,
in contrast with many other systems, the phase space is bounded (the
point vortex model has no quantity analogous to kinetic energy in
particle systems, allowing the system to explore higher and higher
energy with increasing entropy). The same argument also holds for
the 2D Euler equations with continuous vorticity fields, so one expects
negative temperatures to exist for some ranges of energy. In this
section we show that, in the case of a doubly-periodic domain, the
inverse temperature of the 2D Euler equations is actually always negative. Note that the proof has first been established by A. Mikelic \protect\cite{Mikelic_Robert_1998}. 

From the mean-field variational problem we can prove that for any
vorticity distribution there exist functions $f$ such that

\begin{equation}
\omega=\Delta\psi=f(\beta\psi).\label{eq:neg}
\end{equation}
Moreover, it can be proven that $f'(x)>0$ (see for instance Ref. \protect\cite{bouchet_sommeria1}). 

Multiplying the above equation by $\Delta\psi$ and integrating by
parts gives

\begin{equation}
\beta=-\frac{\int_{\mathscr{D}}d^{2}\mathbf{r}\,(\Delta\psi)^{2}}{\int_{\mathscr{D}}d^{2}\mathbf{r}\,(\nabla\psi)^{2}\, f'(\beta\psi)}.
\end{equation}
Using $f'(x)>0$, we conclude that $\beta<0$.\footnote{Note that doubly-periodicity is required to ensure the temperature is always negative. This is due to a partial integration step in the proof.}\\

Let us find the value of $\beta$ in the low energy limit for domain
$\mathscr{D}$. Previous works \protect\cite{bouchet_venaille2} have shown
that in this limit, the expression for the vorticity field for the
parallel flow is given by $\omega=A\cos(2\pi x+\phi)$ (or $\omega=A\cos(2\pi y+\phi)$),
where $A$ is a constant depending on $E$ and $\phi$ is an arbitrary
phase. Furthermore, in the same limit, we may linearize Eq. (\protect\ref{eq:neg})
and find $\omega\simeq\beta\psi$. From $\omega=A\cos(2\pi x+\phi)=\Delta\psi$
we compute $\psi=-(4\pi^{2})^{-1}\omega$. Together with $\omega\simeq\beta\psi$
we find that $\beta=-4\pi^{2}$, which is indeed negative for all
energies.

\subsubsection{Numerical results}

We now confront the analytical results with numerical computations.
In order to do so, we must first ensure that the system is in equilibrium.
We can test this by computing the mean-field entropy. We recall the
entropy of a coarse-grained macrostate $p^{N}$, see Eq. (\protect\ref{eq:specialdeviation}).
The mean-field entropy is computed for the two-level distribution
($K=2,\sigma_{+}=1,\sigma_{-}=-1$) with parameter values $N=256$
and $n=3$, and energies ranging from $E_{0}=0.01E_{max}$ to $E_{0}=0.99E_{max}$.
Here, $E_{max}$ is the maximum possible value of energy the two-level
system can obtain. We will $\textbf{compute}$ $E_{max}$ below. We take $\Delta E=0.01E_{0}$
but note that its value is not of high importance since the actual
energy fluctuations at this resolution are much lower than $\Delta E$.

\begin{figure}[H]

\begin{centering}
\subfloat[]{\begin{centering}
\includegraphics[scale=0.5]{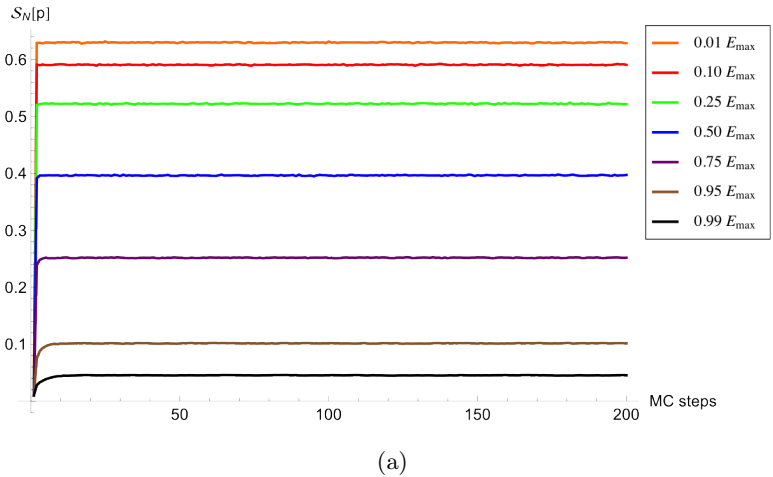}
\par\end{centering}

}\hfill{}\subfloat[]{\begin{centering}
\includegraphics[scale=0.5]{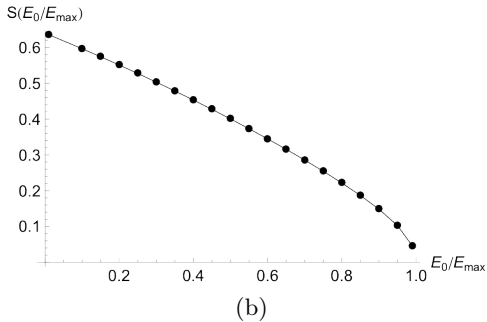}
\par\end{centering}

}\caption{\label{fig:mfe} (a) Mean-field entropy as a function of Monte-Carlo
steps for a two-level vorticity distribution at a resolution of $N=256$.
Equilibrium is reached within fifty Monte-Carlo steps for any energy
level. (b) Mean-field entropy as a function of $E_{0}/E_{max}$. The
numerical fluctuations are smaller than the size of the balls.}

\par\end{centering}

\end{figure}

The result is depicted in Fig. \protect\ref{fig:mfe}(a). Equilibrium is reached
quickly for all energy levels. To be on the safe side, we take an
equilibration time of one-hundred Monte-Carlo steps. 
We define a Monte-Carlo step as $M$ accepted iterations of the Creutz
algorithm. With this definition, the size of a state $x$ and the
Monte-Carlo step scale linearly, such that after $K$ Monte-Carlo
steps each value $x_{i}$ has been changed $K$ times on average.
All results that
follow below have been obtained after the system has reached equilibrium.

$\textit{Two-level vorticity distribution}$. For the 2D Euler equations
and for a given vorticity distribution, energy maxima correspond to
segregated states. For example, when the left half of the vorticity
field contains only values of $\sigma_{+}=1$ and the right half consists
solely values of $\sigma_{-1}=-1$ we have $E_{0}=E_{max}$. 

We calculate $E_{max}$ analytically. We have $d^{2}\psi/dx^{2}=-1$
for $0\leq x<\frac{1}{2}$ and $d^{2}\psi/d^{2}x=+1$ for $\frac{1}{2}\leq x<1$.
The general solutions to these two differential equations on their
respective domains are denoted by $\psi_{-1}$ and $\psi_{1}$. These
equations are complemented with the boundary conditions $\psi_{-1}(0)=\psi_{1}(1)=0$,
$\psi_{-1}(\frac{1}{2})=\psi_{1}(\frac{1}{2})$, and $\psi'_{-1}(\frac{1}{2})=\psi'_{1}(\frac{1}{2})$.
We find $\psi_{-1}(x)=-\frac{x^{2}}{2}+\frac{1}{4}x$ and $\psi_{1}(x)=\frac{x^{2}}{2}-\frac{3}{4}x+\frac{1}{4}$.
The maximum energy is now easily calculated: $E_{max}=\frac{1}{96}.$
The numerical value of $E_{max}=0.0104$ corresponds well to this
theoretical value. 

\begin{figure}[H]
\subfloat[]{\centering{}\includegraphics[scale=0.6]{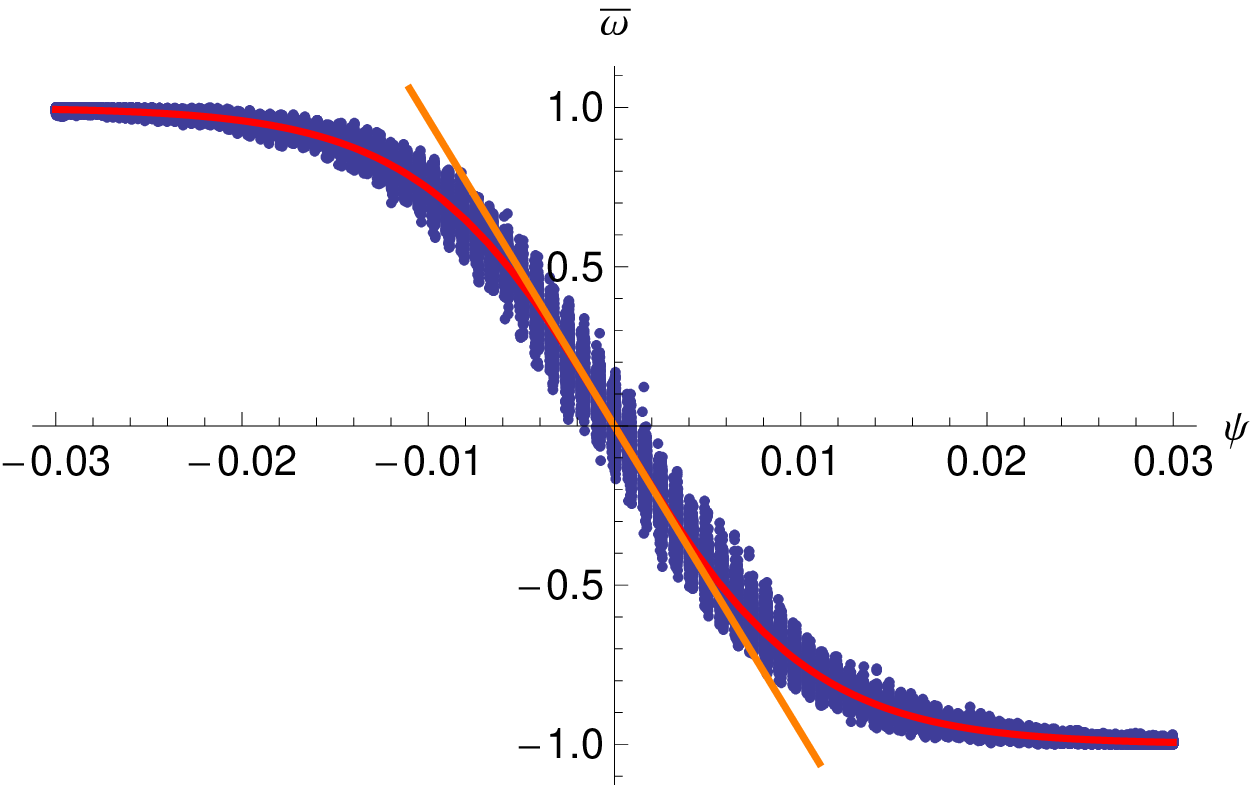}}\hfill{}\subfloat[]{\begin{centering}
\includegraphics[scale=0.5]{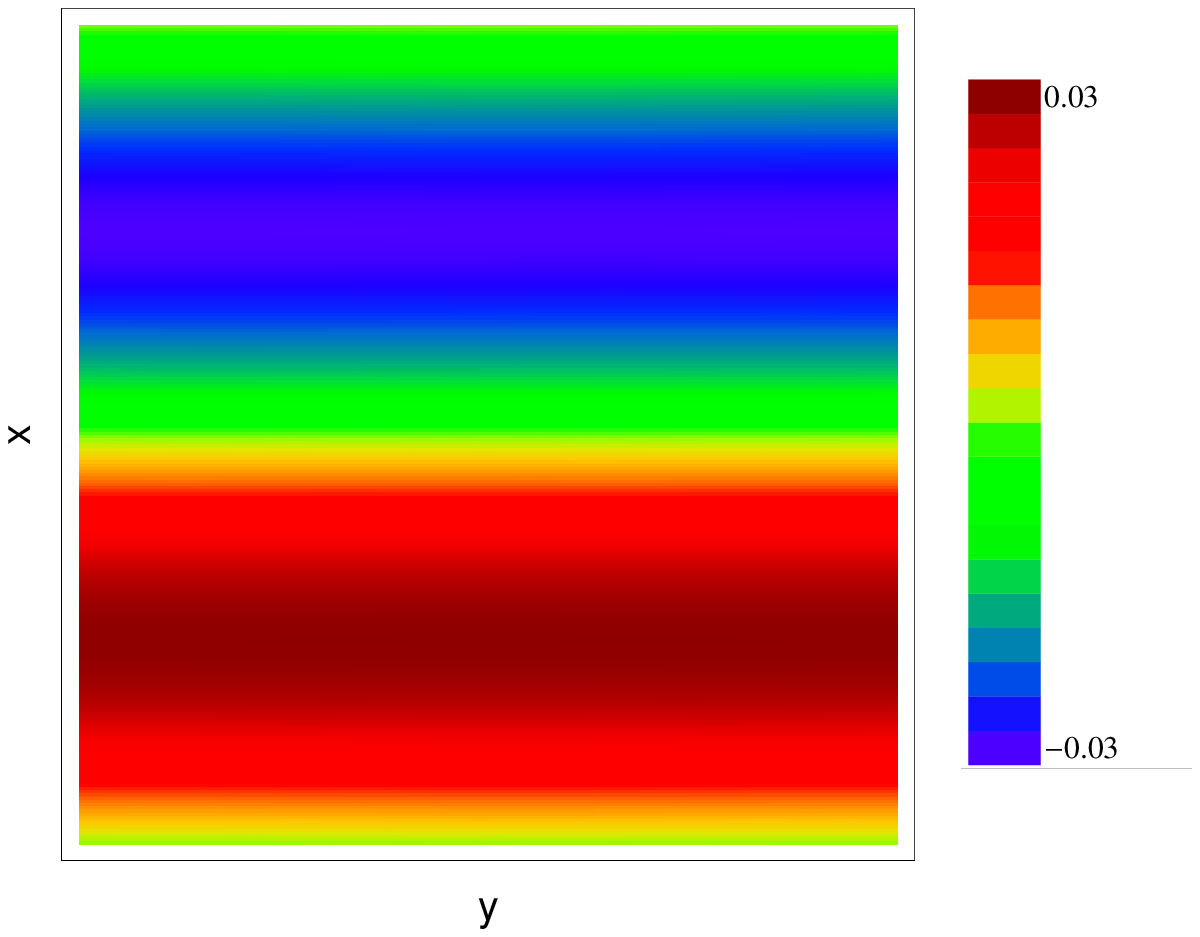}
\par\end{centering}

}\caption{\label{fig:2vcase}Numerical result of a two-level vorticity distribution
with resolution $256\times256$ at energy $E_{0}=0.90E_{max}$ and
coarse-graining parameter $n=3$. (a) Averaged coarse-grained vorticity
versus stream function. An inverse temperature of $\beta=-96.2$ is
found. (b) Averaged stream function field shows a parallel (pure)
flow.}
\end{figure}

We perform numerical simulations at $N=256$ for $E_{0}=0.9E_{Max}$
and an energy tolerance set to $\Delta E=0.01E_{0}$. After reaching
equilibrium, we compute the $p^{N}$ and stream function $\psi^{N}$
by point-wise averaging one-hundred fields which are separated in
time by $N^{2}$ permutations, respectively. The averaged coarse-grained
vorticity field $\bar{\omega}^{N}$ is then computed from $p^{N}$.
Fig. \protect\ref{fig:2vcase}(a) shows the averaged coarse-grained vorticity
$\bar{\omega}^{N}$ versus the averaged stream function $\psi^{N}$.

The data points (blue) are fitted with Eq. (\protect\ref{eq:fit2v}) by tuning
the Lagrange parameter $\beta$. The result (red curve) corresponds
to a temperature of $\beta=-96.2$. The numerical results are clearly
in agreement with the theoretical predictions. In Fig. \protect\ref{fig:2vcase}(b)
the stream function field is plotted, showing a unidirectional flow.

The fluctuations in $\bar{\omega}^{N}$ visible in Fig. \protect\ref{fig:2vcase}(a)
are a result of the coarse graining of $\omega$ (or equivalently,
in $p$). The level of fluctuations in $p^{N}$ is of $o(1/\sqrt{n^{2}})=o(1/n)$.
This amounts to fluctuations of about $0.3$ for $n=3$. This is close
to the observed fluctuations near the center ($\psi=0$), see Fig.
\protect\ref{fig:2vcase}(a). \\

$\textit{Three-level vorticity distribution.}$ Although determining
$E_{max}$ for the two-level case was possible due to the simple geometry,
this is less trivial for the three-level case. Instead of trying to
compute $E_{max}$ analytically, we determine it numerically. We let
the algorithm run for some time and accept only moves which increase
the energy. We find that the system converges towards a maximum energy
of $E_{max}=0.005208$. 

We now show results for a three-level distribution with $N=256$,
$E_{0}=0.9E_{max}$ and an energy tolerance of $\Delta E=0.01E_{0}$.
We follow the same averaging procedure as in the two-level case, see
above. 

\begin{figure}[H]
\subfloat[]{\begin{centering}
\includegraphics[scale=0.6]{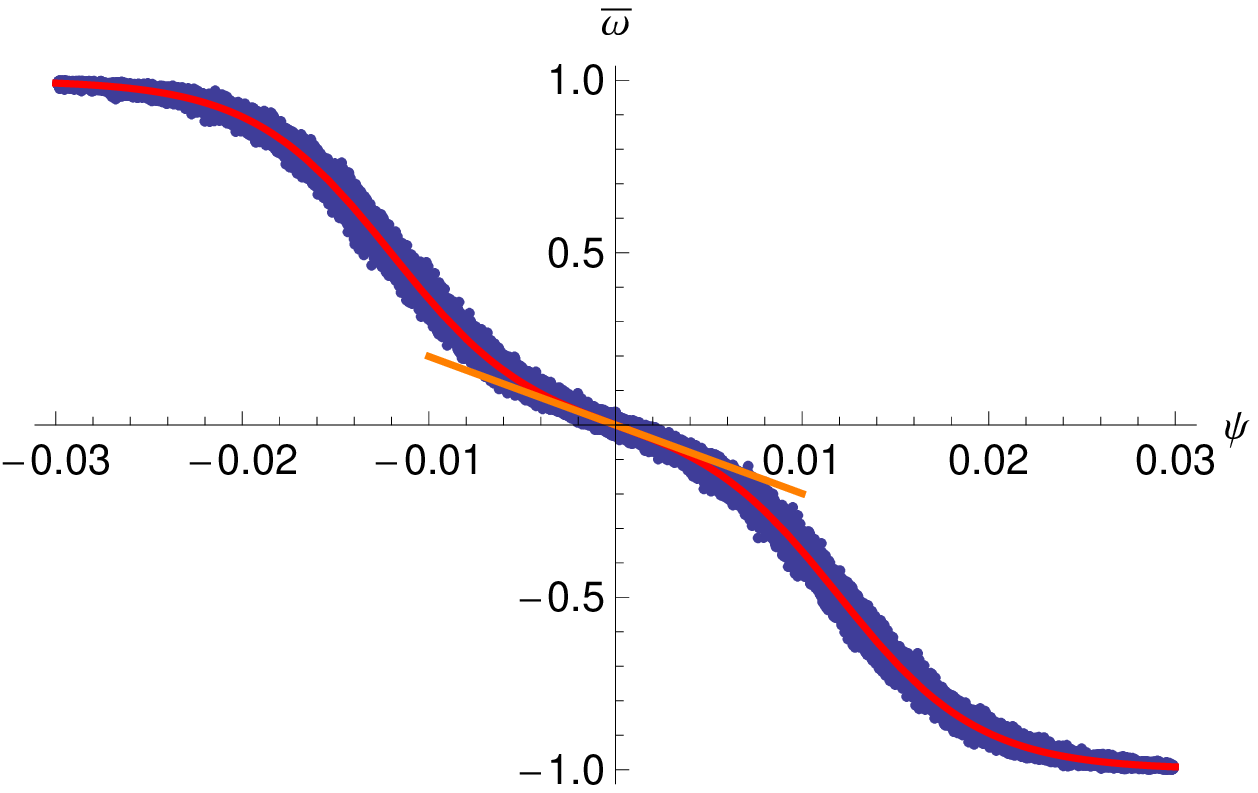}
\par\end{centering}

}\hfill{}\subfloat[]{\begin{centering}
\includegraphics[scale=0.5]{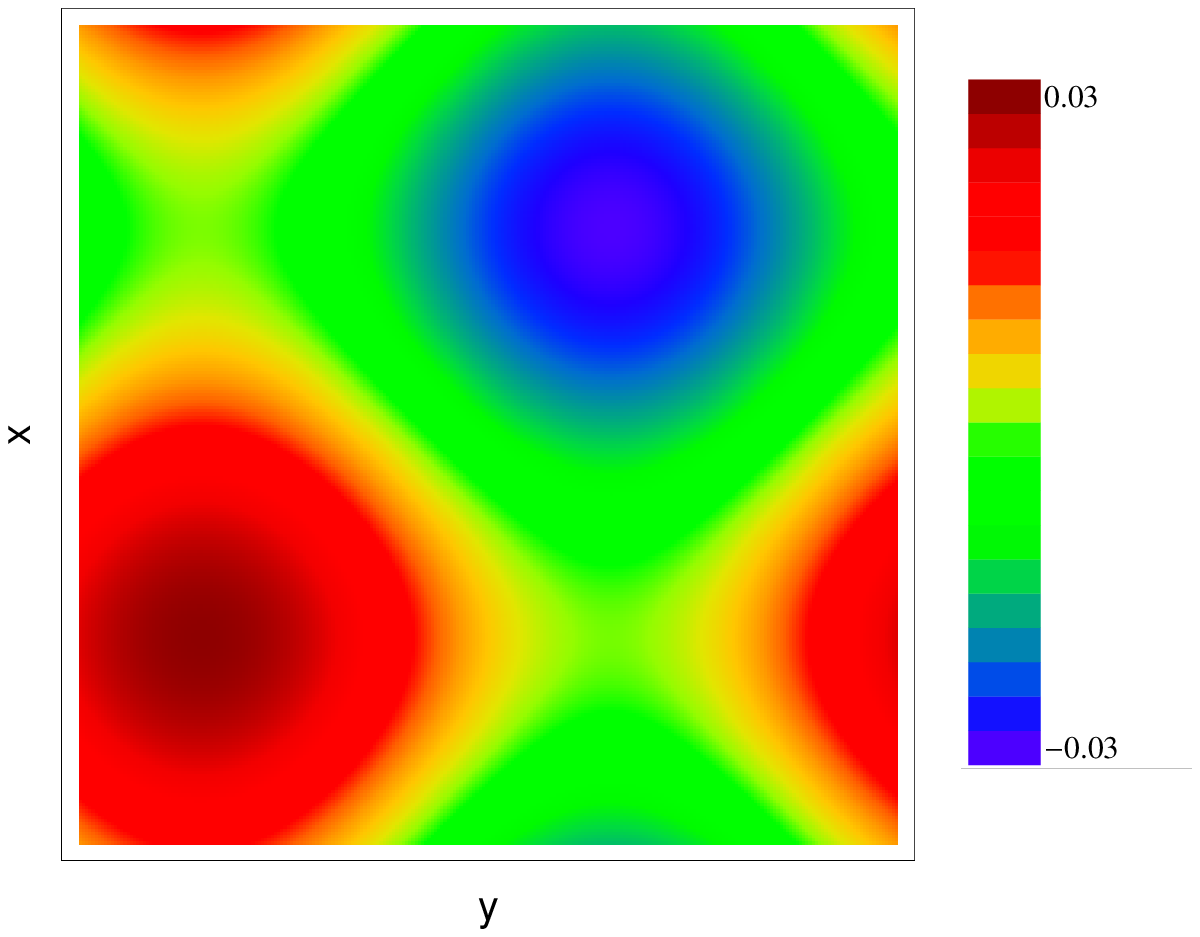}
\par\end{centering}

}\caption{\label{fig:3vcase}Numerical result of a three-level vorticity distribution
with resolution $256\times256$ at energy $E_{0}=0.9E_{max}$ and
coarse-graining parameter $n=3$. (a) Averaged coarse-grained vorticity
versus stream function. Fitting parameters are $\beta=-252.1$ and
$\mu=0.0969$. (b) The averaged stream function field shows a dipole
flow.}
\end{figure}

In Fig. \protect\ref{fig:3vcase}(a) the averaged coarse-grained vorticity
$\bar{\omega}^{N}$ is plotted against the averaged stream function
$\psi^{N}$. The data points (blue) are fitted with Eq. (\protect\ref{eq:fit3v}),
see Fig. \protect\ref{fig:3vcase} for more details. This is again in agreement
with theory. In Fig. \protect\ref{fig:3vcase}(b) the averaged stream function
field is plotted, showing a dipole flow.

\subsection{Inverse temperature computation}

The 2D Euler equations have
the property of negative temperature as discussed in Section \protect\ref{sub:Negative-temperature-in}.

In this section we present three different methods to compute the
inverse temperature of the two-level vorticity system. The first method
is a direct measure of energy fluctuations, as discussed in Section
\protect\ref{sub:Temperature-Creutz}. It is independent of any assumption
(mean-field, large $N$, etc). It is therefore a good method to verify
the two other methods which rely on mean-field approximations. The
second method, studied in Section \protect\ref{sub:Temperature-mfe}, uses
the mean-field equation (\protect\ref{eq:fit2v}) and a fit to compute $\beta$.
The third method, proposed in Section \protect\ref{sub:Temperature-from-mfentro},
uses the mean-field approximation of the entropy (\protect\ref{eq:entropy5})
to find $\beta$ from the relation $\beta=dS/dE$.

We show that all three methods give similar results. Furthermore,
we find negative temperatures for all energies, in agreement with
our finding in Section \protect\ref{sub:Negative-temperature-in}. 

Lastly, in the high energy limit, we find that the inverse temperature
tends to $\beta=-\infty$. In the low energy limit, we find $\beta=-39.74$,
which is close to the theoretically predicted value of $\beta=-4\pi^{2}$
for our domain $\mathscr{D}$, see previous section.

\subsubsection{Temperature from the energy distribution in the Creutz algorithm\label{sub:Temperature-Creutz}}

This method makes use of the theory discussed in Section \protect\ref{sub:Temperature-computation-from}
and is applied to the two-level case. 

\begin{figure}[H]

\subfloat[]{\begin{centering}
\includegraphics[scale=0.6]{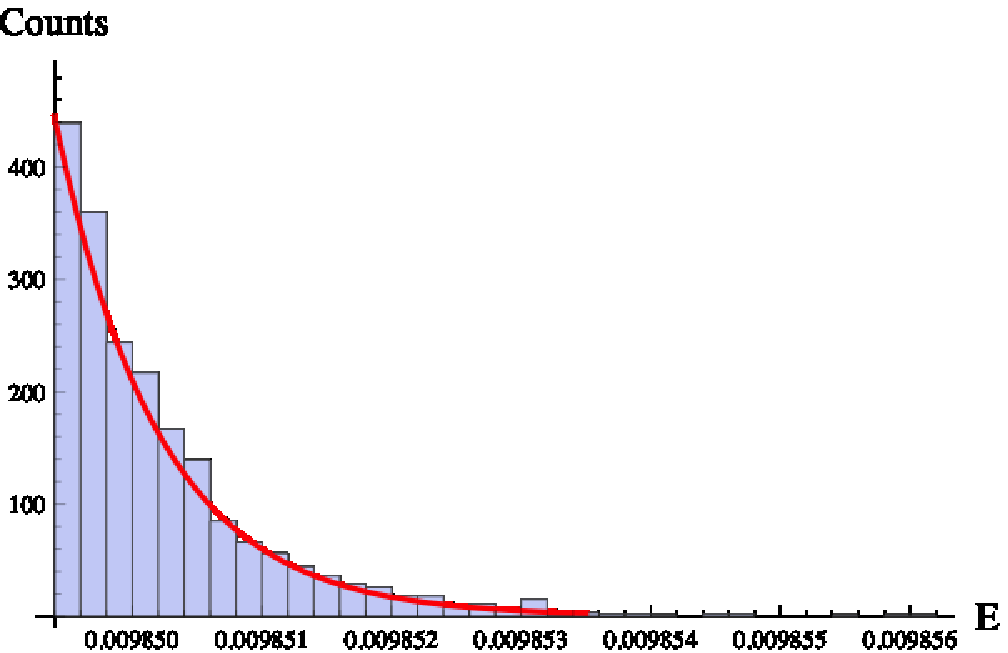}
\par\end{centering}

}\hfill{}\subfloat[]{\begin{centering}
\includegraphics[scale=0.7]{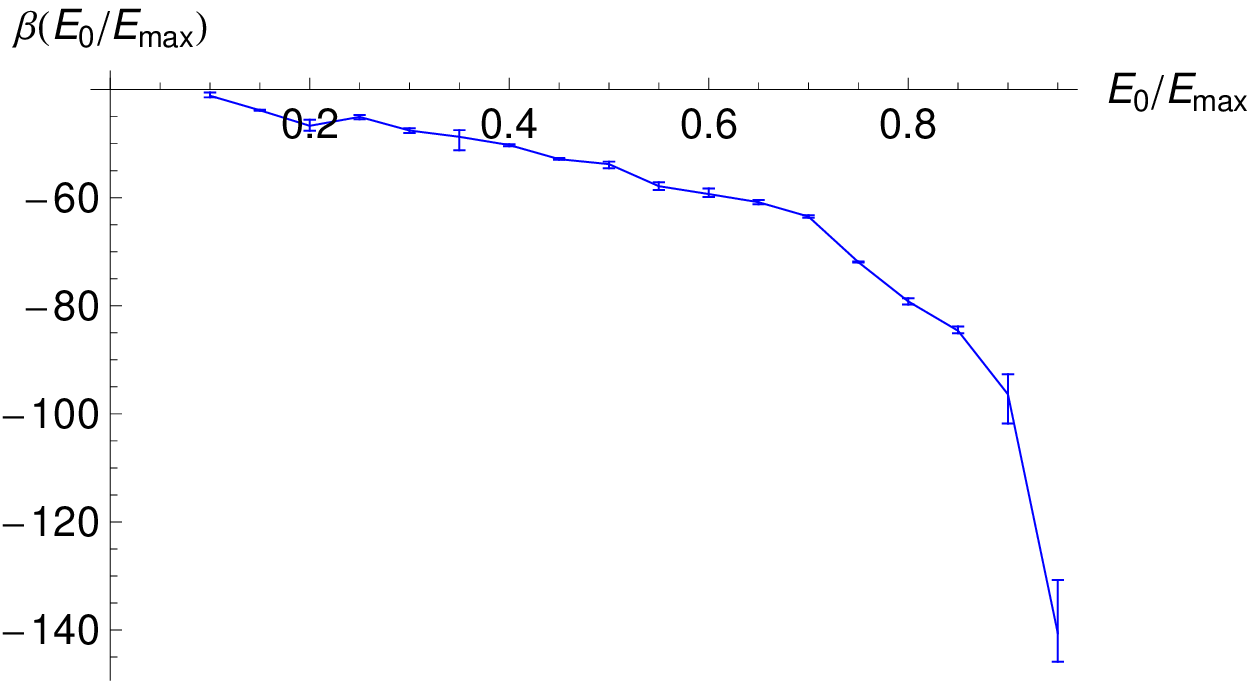}
\par\end{centering}

}\caption{\label{fig:distribution}Two-level vorticity distribution with $N=256$.
(a) Energy distribution after reaching equilibrium. The energy distribution
is proportional to $\Omega_{N}(E)$. The distribution follows an exponential
law (Eq. (\protect\ref{eq:betadistr})), see red curve. This gives access
to the inverse temperature $\beta$. (b) The inverse temperature,
obtained using the method of (a), is plotted for several energies
$E_{0}$. The error bars show the standard error of the estimate of
$\beta$ obtained via the fitting procedure. }

\end{figure}

After equilibrium is reached, the system's energy is computed ten
times per Monte-Carlo step for a total of one-hundred Monte-Carlo
steps. The histogram is shown in Fig. \protect\ref{fig:distribution}(a) for
$E_{0}=0.95E_{max}$. The data is fitted with Eq. (\protect\ref{eq:betadistr}),
from which we obtain a value of $\beta$. 

This method is repeated for different values of energy $E_{0}$, such
that we obtain a plot of $\beta$ versus energy, shown in Fig. \protect\ref{fig:distribution}(b).
Notice that the temperature is negative over the full range of energy
values, as expected.

\subsubsection{Temperature from the mean-field equation\label{sub:Temperature-mfe}}

We compute the inverse temperature through the mean-field equation
(\protect\ref{eq:cgomega}) by using the same method discussed in Section
\protect\ref{sub:Mean-field-predictions}, see also Fig. \protect\ref{fig:2vcase}(a).

\begin{figure}[H]
\begin{centering}
\includegraphics{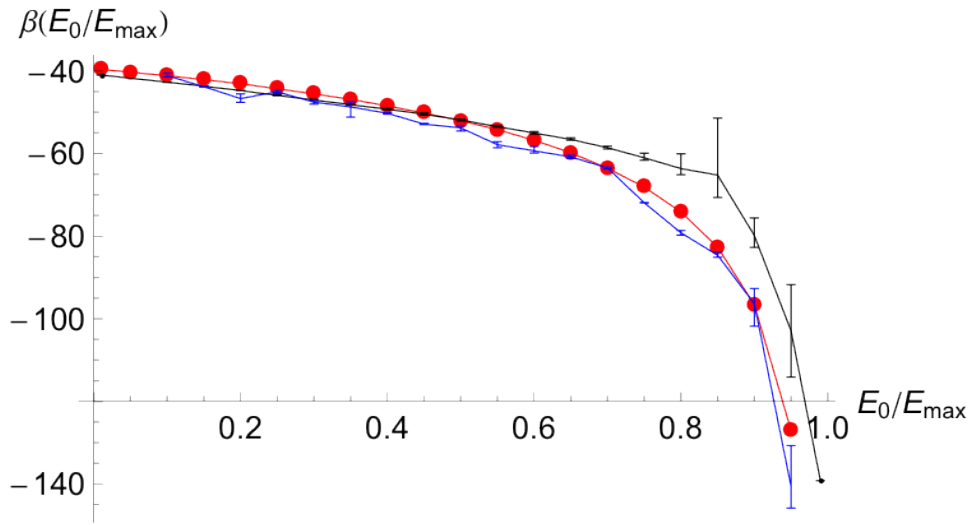}\caption{\label{fig:3methods}Inverse temperature plotted against energy from
the equilibrium energy distribution (blue), the mean-field equation
(red) and the mean-field entropy (black). }

\par\end{centering}

\end{figure}

After the system has reached equilibrium, the average fields $\psi^{N}$
and $\bar{\omega}^{N}$ are computed for several energies $E_{0}$.
We plot $\bar{\omega}^{N}$ versus $\psi^{N}$ for each energy and
fit the result with Eq. (\protect\ref{eq:fit2v}) from which we obtain a negative
temperature $\beta(E)$, see the red curve in Fig. \protect\ref{fig:3methods}. 

\subsubsection{Temperature from the mean-field entropy\label{sub:Temperature-from-mfentro} }

This method relies on the mean-field entropy, see Eq. (\protect\ref{eq:entropy5})
and Section \protect\ref{sub:Mean-field-predictions}.

For each chosen energy $E_{0}$ the algorithm is run until equilibrium
is reached. The simulation is then continued for another fifty Monte-Carlo
steps. We average the coarse-grained entropy and energy over this
interval and repeat this process for energies ranging from $E_{0}=0$
to $E_{0}=E_{max}$. The data is fitted with a polynomial, which can
be derived. The temperature of the system is then found via $\beta(E)=\frac{dS}{dE}(E)$.
The resulting values of $\beta$ for are then plotted against energy,
see black curve in Fig. \protect\ref{fig:3methods}.

From the same figure we can conclude that the temperature is indeed
negative for all values of $E_{0}$. Furthermore, note that all three
methods give very similar results, especially in the low energy limit.
In this limit, statistical mechanics theory predicts a temperature
of $\beta=-4\pi^{2}$, see Section \protect\ref{sub:Negative-temperature-in}.
The values in this limit, shown in the figure, indeed approach this
theoretical value.

\section{\label{sec:Phase-transitions-and}Phase transitions and statistical
ensemble inequivalence}

In this section, we use Creutz's algorithm in order to study the microcanonical
measures for the 2D Euler equations. We are specifically interested
in the study of phase transitions, as they play a major role in the
dynamics of equilibrium and non-equilibrium flows \protect\cite{bouchet_simonnet1,bouchet_venaille2}.
The numerical computations shown in this section are compared with
the low-energy statistical equilibria and phase diagrams analytically
computed in Ref. \protect\cite{bouchet_simonnet1,bouchet_venaille2}. Creutz's
algorithm allows to search beyond the low energy limit. In Section
\protect\ref{sub:Phase-transition-3-levels}, we observe a first-order phase
transition that does not exist in the low-energy limit, for a three-level
vorticity distribution.

\subsection{Summary of theoretical results\label{sub:Summary-of-theoretical} }

We begin with a summary of the theoretical results describing phase
transitions for the 2D Euler equations in the low energy limit \protect\cite{bouchet_venaille2,bouchet_venaille3}.
In a domain with doubly periodic boundary conditions and in the low
energy limit, statistical equilibria are well approximated by largest
scale eigenmodes of the Laplacian 

\begin{equation}
\omega(x,y)\underset{E\rightarrow0}{\sim}A\,\cos(2\pi x+\phi)+B\,\cos(2\pi y+\phi'),\label{eq:vorticity_xy}
\end{equation}
where $A,B$ are constants. 

For a square geometry, there are three possible equilibrium flows.
Two of these equilibria have amplitudes of $(A=0,B\neq0)$ and $(A\neq0,B=0)$,
respectively, and are called pure states, or parallel flows. The last
type of flow, called a symmetric dipole, is the case for which $A=B$.
Symmetric dipoles and of parallel flows have been found numerically
with our algorithm, see previous sections. We proceed now to a more
detailed study of the phase transitions between those equilibria when
energy is changed.

It is shown in Refs. \protect\cite{bouchet_venaille2,bouchet_venaille3} that
the selection of either a dipole or a parallel flow in the low-energy
limit is related to inflection points of the relation between the
coarse-grained vorticity and the stream function. To be more precise,
consider the Taylor expansion 

\begin{equation}
\omega=f(\beta\psi)\,\,\,\mbox{with}\,\,\, f(x)=x+a_{4}x^{3}+o(x^{3}).\label{eq:a4_parameter}
\end{equation}
Using $\beta<0$ (see Section \protect\ref{sub:Negative-temperature-in}), we note that when $a_{4}>0$, the curve $\omega-\psi$ bends upwards
for positive $x$, similar to a hyperbolic sine. When $a_{4}<0$,
then it bends downward for positive $x$, similar to a hyperbolic
tangent. The theory predicts parallel flows when $a_{4}<0$, and dipole
flows when $a_{4}>0$. 

Let us study this criteria in the case of a two-level vorticity distribution.
From the theoretical predictions (Eq. (\protect\ref{eq:fit2v})) we find for
the two-level case with symmetric vorticity distribution ($\alpha=0$)
that $\omega\simeq\beta\psi-\frac{1}{3}\beta^{3}\psi^{3}$, from which
we find that $a_{4}=-\frac{1}{3}<0$, in accordance with the tanh-like
behavior observed (see Fig. \protect\ref{fig:2vcase}(a)). We thus expect
to always observe parallel flows for the two-level case, and no phase
transitions. This is in agreement with the results obtained by using
Creutz's algorithm.\\

The three-level case is more interesting. Linearization of Eq. (\protect\ref{eq:fit3v})
yields $\omega\simeq\frac{\mu}{1+\mu}\beta\psi+\frac{(1-2\mu)\mu}{6(1+\mu)^{2}}\beta^{3}\psi^{3}$.
For $0\leq\mu<1/2$ we find that $a_{4}>0$ and expect to observe
a dipole. This is the case studied in previous section (see Figs.
\protect\ref{fig:3vcase}). Interestingly, in the three-level case, $a_{4}(\mu)=\frac{(1-2\mu)\mu}{6(1+\mu)^{2}}$
is negative for either $\mu>1/2$ or $\mu<0$. This open the possibility
for a phase transition in the three-level case. 

We have not tried to theoretically compute $\mu$ as a function of
the energy as this would be very involved, but we find that a phase
transition actually exists using Creutz's algorithm. We show the result
in next section.\\

The theoretical results \protect\cite{bouchet_venaille2,bouchet_venaille3}
are valid in the low-energy limit. In this limit the theoretical results
indicate that the transition is of second-order (symmetry breaking,
by transition from parallel to dipole flows) and occurs when $a_{4}=0$.
In the following we perform computations both in the low-energy limit
and for large energies. We find that the sign of $a_{4}$ remains
relevant for finding phase transitions. However, we will see that
for large energies the transition can be of first-order.

\subsection{Phase transition in the three-level case\label{sub:Phase-transition-3-levels}}

In order to study phase transition between dipole and parallel flows
we first define an appropriate order parameter. We consider the Fourier
coefficients of the vorticity field $\omega^{N}$, denoted $\hat{\omega}_{\mathbf{k}}$,
where $\mathbf{k}=(k_{x},k_{y})$, such that in Eq. (\protect\ref{eq:vorticity_xy})
$A=\left|\hat{\omega}_{(2\pi,0)}\right|$ and $B=\left|\hat{\omega}_{(0,2\pi)}\right|$.
We define $\omega_{min}=\min\left\{ \left|\hat{\omega}_{(2\pi,0)}\right|,\,\left|\hat{\omega}_{(0,2\pi)}\right|\right\} $
and $\omega_{max}=\max\left\{ \left|\hat{\omega}_{(2\pi,0)}\right|,\,\left|\hat{\omega}_{(0,2\pi)}\right|\right\} $
. Then the order parameter

\begin{equation}
r=\frac{\omega_{min}}{\omega_{max}}
\end{equation}
characterizes the equilibrium state in which the system resides. We
have $0\leq r\leq1$, and for $r=1$, $\omega_{min}=\omega_{max}$,
$A=B$ and we observe a purely symmetric dipole, whereas for $r=0$,
$\omega_{min}=0$, we find that either $A$ or $B$ is equal
to zero, corresponding to either a horizontal or vertical parallel
flow. For an intermediate value of $r$, a mixed state between a parallel
flow and dipole will be observed (which is not a statistical equilibrium
but can be observed transiently).

For $N=128$ we compute the order parameter 256 times per Monte-Carlo
step for energies ranging from $E_{0}=0$ to $E_{0}=E_{max}$ for
the two- and three-level cases, see Fig. \protect\ref{fig:distribution}.
For both cases we found degeneracy at very low energies. Indeed, in
a square box, theory \protect\cite{bouchet_simonnet1,bouchet_venaille2} predicts
a second-order phase transition at $E=0$ as $a_{4}\underset{E\rightarrow0}{\rightarrow}0$.
Then for finite $N$, both parallel and dipole flows are possible
at non-zero but small energies due to finite-size effects. Fig \protect\ref{fig:distribution}(a)
confirms this. It is also expected in Ref. \protect\cite{bouchet_simonnet1,bouchet_venaille2}
that there is no other phase transition for $E>0$ in the case of
two levels of vorticity. Our numerical results agree with this observation.

\begin{figure}[H]
\begin{centering}
\subfloat[]{\begin{centering}
\includegraphics[scale=0.35]{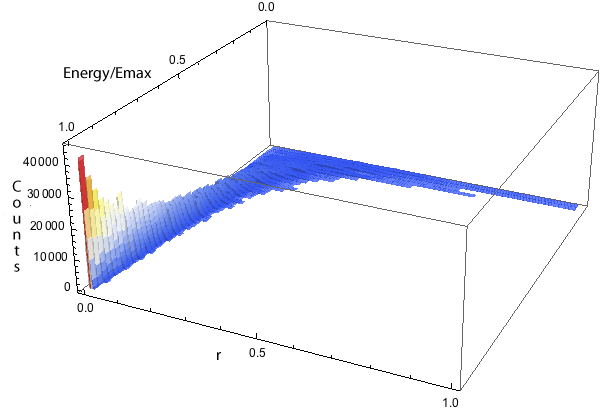}
\par\end{centering}

}\hfill{}\subfloat[]{\begin{centering}
\includegraphics[scale=0.35]{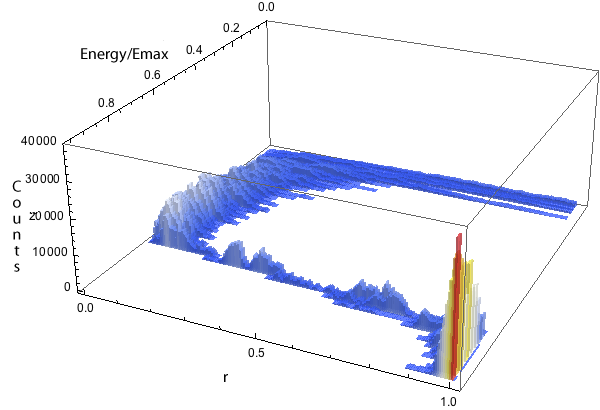}
\par\end{centering}

}\caption{\label{fig:3dhistogram}Order parameter $r$ versus $E/E_{max}=0$
to $E/E_{max}=1.0$ for a two-level (a) and three-level (b) vorticity
distribution. Both distributions show degeneracy at low energies.
Only the three-level case exhibits a phase transition around $E=0.7E_{max}$.}

\par\end{centering}

\end{figure}

For the three-level case, Fig. \protect\ref{fig:3dhistogram}(b) shows that
a phase transition exists around $E_{0}=0.7E_{max}$. \\

\begin{figure}[H]

\centering{}\includegraphics{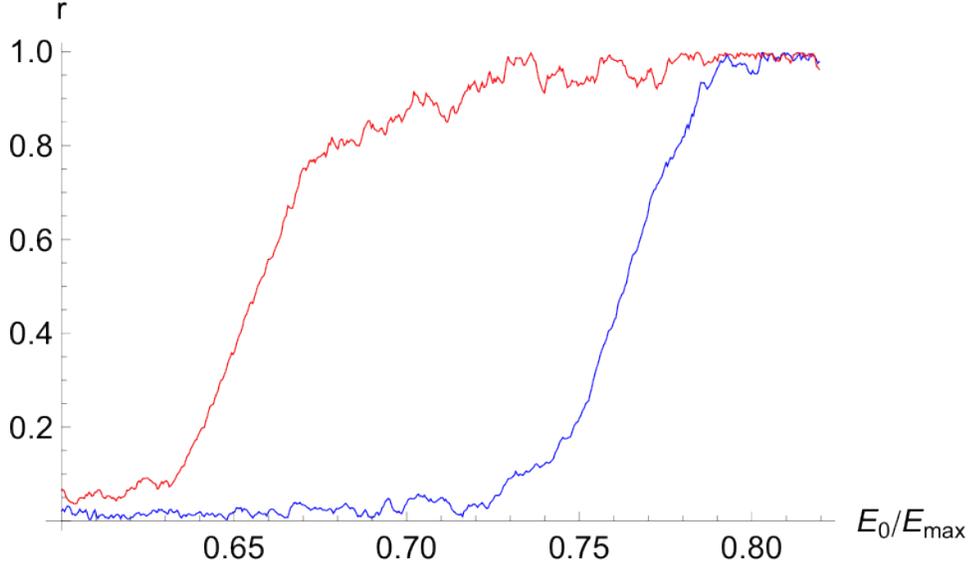}\caption{\label{fig:phasetran}Hysteresis in the transition from parallel flow
to dipole. The plot shows the order parameter versus energy for a
forward (low to high energy) simulation (blue) and a backward (high
to low energy) simulation (red) (resolution N=256).}
\end{figure}

In order to determine the order of the phase transition, we ran Creutz's
algorithm with three levels of vorticity, increasing energy adiabatically
from $E_{m}=0.60E_{max}$ up to $E_{M}=0.82E_{max}$; see Fig. \protect\ref{fig:phasetran}.
We took a resolution of $N=256$ such that fluctuations are small. 

Two simulations were run at this resolution; a forward (low to high
energy) and a backward (high to low energy) simulations. The energy
is adiabatically increased (decreased) with a rate corresponding
to $4.4\times10^{-4}E_{max}$ per Monte-Carlo step, for a total of
500 Monte-Carlo steps. Within each Monte-Carlo step, we compute averages
of the order parameter over 512 realizations, for each of the 500
energy levels. The result is shown in Fig. \protect\ref{fig:phasetran}. We
observe hysteresis behavior, a typical signature of first-order phase
transitions.

Using the standard terminology in describing phase transitions, the
dipole is the ordered phase in the sense that it has lost the translational
symmetry. The parallel flow is the corresponding disordered phase.
An interesting remark is that we observe a transition from the disordered
phase to the ordered one as the energy is increased. This is in contrast
with classical thermodynamics and statistical physics results. For
instance, the first-order solid-liquid transition is a transition
from the ordered phase (solid with broken symmetry) to disordered
phase (liquid) when energy (or temperature) is increased. Similarly,
ferromagnetic-paramagnetic phase transition is a second-order phase
transition from the ordered phase (ferromagnetism, with broken symmetry)
to a disordered phase (paramagnetism) when energy is increased.

This paradox is due to the fact that our system has a negative temperature.
Then entropy decreases with energy. It is thus natural to expect a
transition from the disordered state (high entropy and symmetric)
to the disordered state (low entropy and with broken symmetry) when
energy is increased. This paradoxical property is traced back to the
fact that, in contrast with any other systems, for the 2D Euler equations
the phase space is bounded (there is no equivalent of kinetic energy
allowing the system to explore higher and higher energy with increasing
entropy).\\

We have described a first-order phase transition in the microcanonical
ensemble with hysteretical behavior. In systems with short-range interactions,
microcanonical first-order transitions usually do not exist because of the possibility
of phase coexistence \protect\cite{Bouchet_Barre2005}, whereas it is a generic
feature in systems like the 2D Euler equations which has long-range
interactions. A very general argument \protect\cite{Bouchet_Barre2005} show
that such a first-order phase transition in the microcanonical ensemble
is necessarily associated with a situation of inequivalence between
the microcanonical and canonical ensemble.

\section{\label{sec:Conclusion}Conclusions}

In this paper we have presented a novel numerical method based on
Creutz's algorithm to sample microcanonical measures of hydrodynamical
systems. Although we have only presented numerical results for the
2D Euler model, we stress that this numerical scheme can easily be
generalized to more complex hydrodynamical models such as the axisymmetric
3D Euler equations or the Shallow Water equations.

For the 2D Euler model, we have reproduced the (theoretically) well-known
equilibrium states characterized by parallel and dipole flows. Using
our algorithm, we were able to compute the temperature of the system
in three different ways. One of these approaches allowed the computation
of the temperature without making use of any mean-field assumptions,
and therefore served as a verification tool for the mean-field approximations
made in theoretical predictions. All three methods match very well,
showing consistency between the numerical approach and the mathematical
results, and also proves that the mean-field description is exact
in the large-$N$ limit.

Furthermore, we have found a previously unknown phase transition in
the microcanonical description of the 2D Euler equations. We have
shown that in the energy range where the transition occurs there is
an ensemble inequivalence between the microcanonical and canonical
ensemble. This transition is very interesting from a statistical mechanics
point of view, as it is a transition from an ordered phase with broken
symmetry towards an asymmetric ordered phase when energy is increased.
From a fluid mechanics point of view, it is also very interesting,
as it describes a discontinuous change of the flow topology. A similar
phase transition was already observed in a non-equilibrium framework
for the 2D stochastic Navies-Stokes equations \protect\cite{bouchet_simonnet1}.
This new equilibrium result will probably very useful in explaining
why the 2D-Navier-Stokes non-equilibrium phase transition has the
phenomenology of a first-order phase transition rather than of a second-order
one.

The numerical method we propose is extremely easily implemented and
the generalization to similar models is rather straightforward. We
guess that it will have many applications for theoretical, experimental
and geophysical flows.

\section*{Appendix A Markov chains, detailed balance and invariant distributions}

In this appendix we formalize the notions of Markov chain, detailed
balance, and reversible Markov chain. These definitions are used in
Section \protect\ref{sub:Definition-of-the} and \protect\ref{sub:2D-Euler-algorithm}.\\

$\textit{Markov chain}$ A Markov chain is a (mathematical or physical)
system that undergoes changes from one state to another between a
finite number of states. Such changes are called $\textit{transitions}$
and the probabilities associated with the state changes are called
$\textit{transition probabilities}$. The system is driven by a random
and memoryless process, i.e., the next state in the chain depends
only on the current state and not on the sequence of states that preceded
it. The memoryless property of the system is usually referred to as
the $\textit{Markov property}$. The set of all possible states, called
the $\textit{configuration space}$, and transition probabilities
fully characterizes a Markov chain. 

The configuration space is defined by

\[
X=\left\{ x=(x_{i})_{1\leq i\leq M}\right\} .
\]
A state is thus described by a set of $M$ values: $x=(x_{i})_{1\leq i\leq M}$. 

Formally, a Markov chain $\mathscr{T}$ is a sequence of random variables
$\left\{ y^{l}\in X\right\} _{l\geq0}$ with the Markov property,
where the transition between states are determined by transition probabilities

\[
T(x^{l+1},x^{l},x^{l-1},\ldots,x^{0})=T(x^{l+1},x^{l})=T(x,x').
\]
It gives the probability to go from a state $y^{l}=x'$ to the state
$y^{l+1}=x$. We call a sequence $\left\{ z^{l}\right\} _{l\geq0}$
a $\textit{realization}$ of the Markov chain $\mathscr{T}$. Note
that $i\in\mathbb{N}$ is the index for components of $x$, while
$l\in\mathbb{N}$ is the index for the position in the Markov chain.
Furthermore, we denote $P^{l}(x')$ as the probability for $y^{l}=x'$.
The probability to observe the system in state $y^{l+1}=x$ is then
given by $P^{l+1}(x)=\sum_{x'\in X}T(x,x')P^{l}(x')$. The $\textit{stationary distribution}$,
denoted $P^{\infty}$, is defined such that $P^{\infty}(x)=\sum_{x'\in X}T(x,x')P^{\infty}(x')$.
The stationary distribution is thus invariant under $T$. \\

$\textit{Detailed balance}$ A Markov chain $\mathscr{T}$ is said
to be $\textit{reversible}$ with respect to the distribution $P^{\infty}(x)$
if 

\begin{equation}
\forall x,x'\in X:\;\; P^{\infty}(x)T(x,x')=P^{\infty}(x')T(x',x).\label{eq:detailed}
\end{equation}
This condition is also known as the $\textit{detailed balance}$ condition.
Summing Eq. (\protect\ref{eq:detailed}) over $x$ gives

\begin{equation}
\forall x'\in X:\;\sum_{x\in X}P^{\infty}(x')T(x,x')=\sum_{x\in X}P^{\infty}(x')T(x',x)=P^{\infty}\sum_{x\in X}T(x',x)=P^{\infty}(x').
\end{equation}
Hence, for reversible Markov chains, $P^{\infty}$ is always a steady-state
(stationary) distribution of $\mathscr{T}$. In the case where $P^{\infty}(x)$
is uniform over $X$, i.e. $P^{\infty}(x)=\text{cste}$, the detailed
balance condition reduces to

\begin{equation}
\forall x,x'\in X:\;\; T(x,x')=T(x',x).
\end{equation}
We thus find that if a Markov chain obeys detailed balance, there
exists an invariant (stationary) distribution $P^{\infty}$ over $X$.

\section*{Appendix B A wrong way to define and use the Creutz algorithm}

A common error in using the Creutz algorithm is to define $y^{l+1}$
as the first accepted move after $y^{l}$ with the condition $E_{0}\leq\mathscr{E}_{N}(y^{l+1})\leq E_{0}+\Delta E$,
hereby discarding any unaccepted state in the expectation value of
an observable A, c.f. Eq. (\protect\ref{eq:sumobs}). We consider the Markov
chain that corresponds to this wrong procedure and show that it does
not verify detailed balance.

Let us define more precisely the wrong algorithm. Given any configuration,
we randomly pick a new configuration $y^{l+1}$ with probability $T(x,x')$.
If $E_{0}\leq\mathscr{E}_{N}(y^{l+1})\leq E_{0}+\Delta E$ we accept
the move. If the condition is not satisfied we keep picking at random
new values for $y^{l+1}$ until the condition $E_{0}\leq\mathscr{E}_{N}(x)\leq E_{0}+\Delta E$
is fulfilled and then accept the move. This defines a Markov chain
$\mathscr{R}$ with transition probability

\begin{equation}
R(x,x')=C(x')T(x,x'),
\end{equation}
where $C(x')=\sum_{x\in\Gamma_{N}(E,\Delta E)}T(x,x')$ is called
the acceptance ratio (depending on $x'$ only). 

The detailed balance condition would require $\forall x,x'\in\Gamma_{N}(E,\Delta E):\; R(x,x')=R(x',x)$.
This would only hold when $C(x')=C(x)$, but there is no reason for
this to be true in general. 

We now recall how to properly empirically sample an observable $A$.
The expectation value of observable $A$ is computed through

\begin{equation}
<A(y)>=\lim_{L\rightarrow\infty}\frac{1}{L}\sum_{l=1}^{L}A(y^{l}).\label{eq:sumobs-1}
\end{equation}
In using the above defined wrong Creutz algorithm, the expectation
value of $A$ is calculated over $\mathscr{R}$. Since the detailed
balance condition does not hold for this Markov chain, one does not
sample a stationary measure.

\section*{Acknowledgements}

This research has been supported through the ANR program STATOCEAN
(ANR-09-SYSC-014) and ANR STOSYMAP (ANR-2011-BS01-015). Numerical
results have obtained with the PSMN platform in ENS-Lyon.

\bibliographystyle{unsrt}
\bibliography{references}

\begin{thebibliography}{10}

\bibitem{bouchet_venaille2}
Freddy Bouchet and Antoine Venaille.
\newblock Statistical mechanics of two-dimensional and geophysical flows.
\newblock {\em Physics Reports}, 515(5):227 -- 295, 2012.

\bibitem{Lim}
C.~{Lim} and J.~{Nebus}.
\newblock {\em {Vorticity, Statistical Mechanics, and Monte Carlo Simulation}}.
\newblock Springer, 2007.

\bibitem{Majda_Wang_Book_Geophysique_Stat}
A.~J. {Majda} and X.~{Wang}.
\newblock {\em {Nonlinear Dynamics and Statistical Theories for Basic
  Geophysical Flows}}.
\newblock Cambridge University Press, 2006.

\bibitem{Lim_Ding_Nebus_2010}
C.~{Lim}, X.~{Ding}, and J.~{Nebus}.
\newblock {\em {Vortex dynamics, Statistical mechanics and Planetary
  atmospheres}}.
\newblock World Scientific, 2010.

\bibitem{bouchet_sommeria1}
F.~Bouchet and J.~Sommeria.
\newblock Emergence of intense jets and jupiter's great red spot as maximum
  entropy structures.
\newblock {\em J. Fluid. Mech.}, 464:165--207, 2002.

\bibitem{bouchet_venaille1}
F.~Bouchet and A.~Venaille.
\newblock Ocean rings and jets as statistical equilibrium states.
\newblock {\em J. Phys. Oceanography}, in press, 2011.

\bibitem{bouchet_simonnet1}
F.~Bouchet and E.~Simonnet.
\newblock Random changes of flow topology in two-dimensional and geophysical
  turbulence.
\newblock {\em Phys. Rev. Lett.}, 102, 2009.

\bibitem{Bouchet_Barre2005}
F.~{Bouchet} and J.~{Barre}.
\newblock Classification of phase transitions and ensemble in equivalence in
  systems with long range interactions.
\newblock {\em Journal of Statistical Physics}, 118, 2005.

\bibitem{Bouchet_Gupta_Mukamel2010}
F.~{Bouchet}, S.~{Gupta}, and D.~{Mukamel}.
\newblock Thermodynamics and dynamics of systems with long-range interactions.
\newblock {\em Physica A: Special Issue FPSP XII 389, 4389}, 2010.

\bibitem{Campa_Dauxois_Ruffo2009}
A.~{Campa}, T.~{Dauxois}, and S.~{Ruffo}.
\newblock Statistical mechanics and dynamics of solvable models with long-range
  interactions.
\newblock {\em Physics Reports}, 480:57--159, 2009.

\bibitem{Chavanis_2006IJMPB_Revue_Auto_Gravitant}
P.~{Chavanis}.
\newblock Phase transitions in self-gravitating systems.
\newblock {\em International Journal of Modern Physics B}, 20:3113--3198, 2006.

\bibitem{EllisHavenTurkington:2000_Inequivalence}
R.~S. {Ellis}, K.~{Haven}, and B.~{Turkington}.
\newblock {Large Deviation Principles and Complete Equivalence and
  Nonequivalence Results for Pure and Mixed Ensembles}.
\newblock {\em J. Stat. Phys.}, 101:999, 2000.

\bibitem{Bouchet_PhysicaD_2008}
F.~{Bouchet}.
\newblock Simpler variation problems for statistical equilibria of the 2d euler
  equation and other systems with long-range interactions.
\newblock {\em Physica D}, 237:1976--1981, 2008.

\bibitem{creutz}
M.~Creutz.
\newblock Microcanonical monte carlo simulation.
\newblock {\em Phys. Rev. Lett.}, 50, 1983.

\bibitem{Mukamel-Ruffo_Schreiber_2005PhRvL}
D.~{Mukamel}, S.~{Ruffo}, and N.~{Schreiber}.
\newblock {Breaking of Ergodicity and Long Relaxation Times in Systems with
  Long-Range Interactions}.
\newblock {\em Phys. Rev. Lett.}, 95(24):240604, 2005.

\bibitem{Gupta_Mukamel2010}
S.~{Gupta} and D.~{Mukamel}.
\newblock Relaxation dynamics of stochastic long-range interacting systems.
\newblock {\em Journal of Statistical Mechanics}, 2010.

\bibitem{robert_michel}
J.~Michel and R.~Robert.
\newblock Large deviations for young measures and statistical mechanics of
  infinite-dimensional dynamical systems with conservation laws.
\newblock {\em Communications in Mathematical Physics}, 159:195--215, 1994.

\bibitem{Boucher_Ellis_Turkington_2000_JSP}
C.~{Boucher}, R.~S. {Ellis}, and B.~{Turkington}.
\newblock {Derivation of maximum entropy principles in two-dimensional
  turbulence via large deviations}.
\newblock {\em J. Stat. Phys.}, 98(5-6):1235, 2000.

\bibitem{Bouchet_Corvellec2010}
F.~{Bouchet} and M.~{Corvellec}.
\newblock Invariant measures of the 2d euler and vlasov equations.
\newblock {\em Journal of Statistical Mechanics}, 2010.

\bibitem{Salmon2010}
R.~{Salmon}.
\newblock The shape of the main thermocline, revisited.
\newblock {\em Journal of Marine Research}, 68:541--568, 2010.

\bibitem{Dubinkina_Frank_2010JCoPh}
S.~{Dubinkina} and J.~{Frank}.
\newblock {Statistical relevance of vorticity conservation in the Hamiltonian
  particle-mesh method}.
\newblock {\em Journal of Computational Physics}, 229:2634--2648, April 2010.

\bibitem{turkington1}
B.~Turkington and N.~Whitaker.
\newblock Statistical equilibrium computations of coherent structures in
  turbulent shear layers.
\newblock {\em J. Sci. Comput.}, 17:1414--1433, 1996.

\bibitem{turkington2}
N.~Whitaker and B.~Turkington.
\newblock Maximum entropy states for rotating vortex patches.
\newblock {\em Physics of Fluids}, 6:3963--3973, 1998.

\bibitem{robert_sommeria}
R.~Robert and J.~Sommeria.
\newblock Relaxation towards a statistical equilibrium state in two-dimensional
  perfect fluid dynamics.
\newblock {\em Phys. Rev. Lett.}, 69:2776--2779, 1992.

\bibitem{thess_sommeria}
A.~Thess and J.~Sommeria.
\newblock Intertial organization of two-dimensional turbulent vortex sheet.
\newblock {\em Phys. Fluids}, 6:2417--2429, 1994.

\bibitem{Cohen_Mukamel_2012}
O.~{Cohen} and D.~{Mukamel}.
\newblock Ensemble inequivalence: Landau theory and the abc model.
\newblock {\em preprint}, 2012.

\bibitem{esm5}
R.~Salmon.
\newblock {\em Lectures on Geophysical Fluid Dynamics}.
\newblock Oxford University Press, 1998.

\bibitem{bouchet_thalabard}
F.~Bouchet, P.J. Morisson, S.~Thalabard, and O.~Zaboronksi.
\newblock On liouville's theorem for non-canonical hamiltonian systems with
  application to fluid and plasma dynamics. a special class of stationary flows
  for the two-dimensional euler equations: A statistical mechanics description.
\newblock {\em Preprint, to be submitted soon}, 2012.

\bibitem{robert3}
R.~Robert.
\newblock On the statistical mechanics of the 2d euler equations.
\newblock {\em Communications in Mathematical Physics}, 212:245--256, 2000.

\bibitem{Newman_Barkema_1999}
M.~{Newman} and G.~{Barkema}.
\newblock {\em {Monte Carlo Methods in Statistical Physics}}.
\newblock Clarendon Press, 1999.

\bibitem{juttner_thess}
B.~Juttner and A.~Thess.
\newblock On the symmetry of self-organized structures in two-dimensional
  turbulence.
\newblock {\em Phys. Fluids}, 7:2108--2110, 1995.

\bibitem{onsager1949}
L.~Onsager.
\newblock Statistical hydrodynamics.
\newblock {\em Il Nuovo Cimento (1943-1954)}, 6:279--287, 1949.

\bibitem{Mikelic_Robert_1998}
A.~{Mikelic} and R.~{Robert}.
\newblock On the equations describing a relaxation toward a statistical
  equilibrium state in the two-dimensional perfect fluid dynamics.
\newblock {\em SIAM J. Math. Anal.}, 29:1238--1255, 1998.

\bibitem{bouchet_venaille3}
A.~Venaille and F.~Bouchet.
\newblock Statistical ensemble inequivalence and bicritical point for
  two-dimensional flows and geophysical flows.
\newblock {\em Phys. Rev. Lett. 102}, 10:104501--+, 2009.

\end{thebibliography}

\end{document}